\documentstyle[pra,aps,twocolumn,floats,psfig]{revtex}
\newcommand{\mrm}{\rm}

\begin{document}

%\special{papersize=8.5in,11in}

\draft

\wideabs{

\title{Calculations of collisions between cold alkaline earth atoms in
a weak laser field}

\author{Mette Machholm}

\address{Department of Computational Science, The National University
of Singapore, Singapore 119260}

\author{Paul S. Julienne}

\address{National Institute for Standards and Technology, 100 Bureau
Drive, Stop 8423, Gaithersburg, MD 20899-8423}

\author{Kalle-Antti Suominen}

\address{Department of Applied Physics, University of Turku,
FIN-20014 Turun yliopisto, Finland\\
Helsinki Institute of Physics, PL 64, FIN-00014 Helsingin yliopisto,
Finland\\
{\O}rsted Laboratory, NBIfAFG, University of Copenhagen,
Universitetsparken 5, DK-2100 Copenhagen {\O}, Denmark}

\date{\today}

\maketitle

\begin{abstract}
We calculate the light-induced collisional loss of laser-cooled and
trapped magnesium atoms for detunings up to 50 atomic linewidths to
the red of the $^1$S$_0$-$^1$P$_1$ cooling transition.  We evaluate
loss rate coefficients due to both radiative and nonradiative
state-changing mechanisms for temperatures at and below the Doppler
cooling temperature.  We solve the Schr\"{o}dinger equation with a
complex potential to represent spontaneous decay, but also give
analytic models for various limits.  Vibrational structure due to
molecular photoassociation is present in the trap loss spectrum. 
Relatively broad structure due to absorption to the
Mg$_2$${^1}\Sigma_u$ state occurs for detunings larger than about 10
atomic linewidths.  Much sharper structure, especially evident at low
temperature, occurs even at smaller detunings due to of Mg$_2$$^1\Pi_g$
absorption, which is weakly allowed due to relativistic retardation
corrections to the forbidden dipole transition strength.  We also perform
model studies for the other alkaline earth species Ca, Sr, and Ba and for Yb,
and find similar qualitative behavior as for Mg.
\end{abstract}

}

%\bigskip

\narrowtext

\section{Introduction}

\subsection{Background}

Laser cooling and trapping of neutral atoms has recently opened many
new research areas in atomic physics.  One can cool a gas of neutral
atoms in magneto-optical traps (MOT) down to temperatures of 1 mK and
below, and obtain densities up to 10$^{12}$ atoms/cm$^3$.  Evaporative
cooling methods have allowed the cooling of alkali species to much
lower temperatures below 1 $\mu$K so that Bose-Einstein condensation
(BEC) occurs.  Binary atomic collisions play an important role in the
physics of cold trapped atomic gases, and have been widely
investigated~\cite{Weiner99}.  One of the first cold collisional process to
be studied is the heating and loss of trapped atoms which result from
tuning laser light to near-resonance with the atomic cooling
transition~\cite{Sesko89}.  Here we take near-resonance, or
small-detuning, to mean detuning $\Delta$ up to 50 natural linewidths
to the red of atomic resonance.

Studies of small-detuning trap loss, extensively reviewed by Weiner
{\it et al.}~\cite{Weiner99}, have mainly concentrated on alkali
atoms~\cite{rare}, for which it has been very difficult to develop
quantitative theoretical models to compare with experiment.  This is
because alkali atoms have extensive hyperfine structure, and thus the
number of collision channels is simply too large for accurate
theoretical modeling.  It has even been difficult to estimate the
relative weight of the different possible loss processes.  Although
one can develop simplified models, these are difficult to test
adequately with complex alkali systems.  On the other hand, trap loss
photoassociation spectra in alkali systems for large detuning can be
modeled quite accurately~\cite{Jones96,Burke99,Williams99}.  This is
because the underlying molecular physics of alkali dimer molecules is
well-known, and the spectra are determined by isolated molecular
vibrational-rotational levels, for which the photoassociation line
shapes can be well-characterized even in the presence of hyperfine
structure.  Quantitative analysis of such spectra have permitted the
determination of scattering lengths for ground state
collisions~\cite{Weiner99}. These scattering lengths are critical
parameters for BEC studies.

Alkaline earth cooling and trapping have recently been of considerable
experimental interest.  Trap loss collisions have been studied in a Sr
MOT~\cite{Gallagher99}, and intercombination line cooling of Sr has
resulted in temperatures below 1~$\mu$K and raised the prospects of
BEC of Sr~\cite{Katori99,Katori99b,Ido00,Katori00}.  Ca is of interest
for possible applications as an optical frequency
standard~\cite{Sengstock94,Ruschewitz98,Riehle98,Riehle99,Oates99},
and photoassociation spectroscopy in a Ca MOT has been
reported~\cite{Tiemann00}.  Intercombination line cooling has also
been reported for Yb~\cite{Honda99,Kuwamoto99}, which we have included
in our discussion because of its similarity in structure to alkaline
earth atoms.

Alkaline earth species provide an excellent testing ground for cold
collision theories, especially given the rapidly developing
experimental interest in the subject.  Since the main isotopes of
alkaline earth atoms have no hyperfine structure, the number of
collision channels becomes low enough to allow theoretical
calculations even in the small detuning trap loss regime.
Consequently, this paper presents theoretical predictions for small
detuning trap loss spectra in cold and trapped Mg gas in the presence
of near-resonant weak laser light tuned near the $^1$S$_0\to{^1}$P$_1$
atomic transition, and discusses the nature of similar processes for
Ca, Sr, Ba, and Yb.  This work extends our previous note on Mg trap
loss~\cite{Machholm99} to other species and lower temperatures, and
shows the relation between small detuning trap loss and
photoassociation theory.  It is necessary to include spontaneous
emission in modeling trap loss collision dynamics because of the long
time scale of cold collisions.  In addition, relativistic retardation
effects play a prominent role at small detunings by allowing
transitions to the dipole forbidden ${^1}\Pi_g$ state, which
exhibits resolved vibrational structure, especially at very low
temperature.  Although we treat the long-range molecular interactions
accurately, the potential energy curves and coupling matrix elements
for the dimer molecules in the short-range region of chemical bonding
are not sufficiently well-known to determine all aspects of trap loss.
Therefore, we examine the uncertainties associated with the unknown
phases developed in the short-range region of chemical bonding, and
show which features are robust with respect to such uncertainties and
which must be measured or later determined from improved theory.

\subsection{Trap loss collisions}

Light-induced trap loss takes place as a molecular process.  Two
colliding cold atoms form a quasimolecule, and their motion can be
described in terms of the electronic (Born-Oppenheimer) potentials of
the molecular dimer with light-induced transitions between the
molecular states.  We consider only red detuning, which excites
attractive potentials at long range.  Such potentials support a number
of bound vibrational states.

\begin{figure}[htb]
%\vspace{7cm}
\noindent\centerline{
\psfig{width=80mm,figure=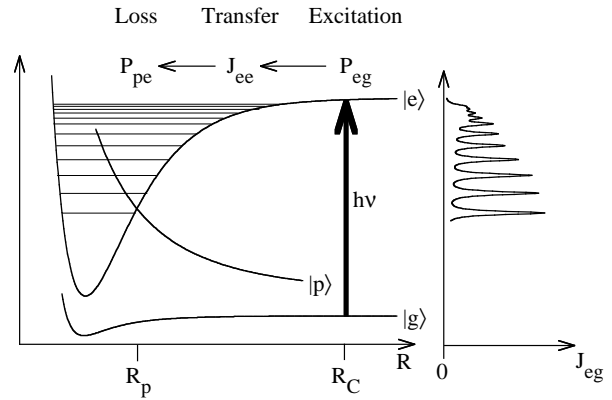}
}
\caption[f1]{Schematic representation of trap loss collisions.
\label{mechanisms}}
\end{figure}

Figure~\ref{mechanisms} schematically indicates the nature of the trap
loss process in a weak radiation field.  An excitation laser with
energy $h\nu$ is tuned a few atomic linewidths below the atomic
transition energy $h\nu_0$.  The ground $|g\rangle$ and excited
$|e\rangle$ quasimolecular electronic states are thus coupled
resonantly by the laser at some long-range Condon radius $R_C$, where
the photon energy matches the difference between the excited and
ground potential energy curves.  The excited state decays to a loss
channel $|p\rangle$ due to interactions at short range.  The fully
quantum mechanical description in Ref.~\cite{Julienne94} shows that
the overall probability $P_{pg}$ of a trap loss collision can be
factored as follows:
\begin{equation}
        P_{pg}=P_{pe}J_{eg} .
        \label{factor1}
\end{equation}
Here $J_{eg}$ represents an excitation-transfer probability which is
proportional to the scattering flux that reaches the short range
region near $R_p$ due to long-range excitation near the Condon point
followed by propagation on the excited state to the short-range
region.  $P_{pe}$ represents the probability of a transition from the
excited state to the loss channel at short range during a single cycle
of oscillation through the short-range region.

Figure~\ref{mechanisms} indicates the qualitative behavior of the
transfer function $J_{eg}$ versus detuning $\Delta$.  At very small
detuning of a few atomic linewidths $J_{eg}$ becomes very small if
there is a high probability of spontaneous emission during transit
from the outer to inner regions.  At sufficiently large detuning
$J_{eg}$ exhibits resonance structure due to the bound vibrational
levels in the excited state potential, where the vibrational period is
much shorter than the decay time~\cite{Flux}.  This is the domain of
photoassociation to individual vibrational levels.
Section~\ref{LimitingCases} below will show simple analytic formulas
for $J_{eg}$ that apply in these two limiting cases of small detuning
with fast decay or isolated photoassociation lines.  These formulas
show how $J_{eg}$ can in turn be factored as
\begin{equation}
        J_{eg}=J_{ee}P_{eg} ,
        \label{factor2}
\end{equation}
where $P_{eg}$ represents the probability of excitation from the
ground to excited state in the outer zone near the Condon point, and
$J_{ee}$ represents an excited state transfer function between the
long-range outer excitation zone and the short-range zone.   The
factorization in Eq.~(\ref{factor2}) is also schematically indicated
in Fig.~\ref{mechanisms}.

Each of the factors in Eqs.~(\ref{factor1}) and (\ref{factor2}) can be
affected by unknown phases associated with the short-range molecular
physics of the dimer molecules.  (1) $P_{eg}$ is sensitive to the
asymptotic phase of the ground state wave function.  (2) Vibrational
resonance structure in $J_{eg}$ is sensitive to the position and
widths of vibrational features.  (3) $P_{pe}$ is sensitive to
St\"{u}ckelberg oscillations in short-range curve crossing
probabilities.  These effects are discussed in detail in Secs.
\ref{GroundState} and \ref{ShortRange}.  The overall effect of such
sensitivities will depend on the temperature and the alkaline earth
species.

There are two possible inner zone loss processes characterized by
$P_{pe}$: the state change (SC) and radiative escape (RE) mechanisms,
both represented schematically by the loss channel $p$ in
Fig.~\ref{mechanisms}.  In the SC process the excited
state couples to another molecular state near a short-range crossing
point $R_p$, and population transfer between them is possible.  The
products of the collision emerge on a state that correlates
asymptotically with other atomic states of lower energy, such as
${^3}$P$+{^1}$S, thereby releasing a large amount of kinetic energy to
the separating atoms.  In RE the excited state can decay
by spontaneous emission after the atoms have been accelerated towards
each other on the excited state potential.  The ground state atoms
then separate with this gain in kinetic energy.  If enough kinetic
energy has been gained to exceed the trap depth, this is a loss
process (early decay after insufficient acceleration only leads to
radiative heating).  Here $R_p$ represents the distance at which the
atoms have received sufficient acceleration to be lost from the trap.
Any emission with $R < R_p$ leads to trap loss.

In this paper we use the fully quantum complex potential method of
Ref.~\cite{Julienne94} and do not resort to semiclassical methods
(with one exception), although semiclassical concepts are often useful
for interpretation.  Our quantum methods are fully capable of
describing the $s$-wave limit and the quantum threshold properties of
the extremely low collision energies near or below the critical
temperature for BEC. We describe the spontaneous emission processes
with a complex potential, and solve the corresponding time-independent
multichannel Schr\"odinger equation in the molecular electronic state
basis.  This takes any vibrational state structure into account
automatically without the need to calculate wave functions or
Franck-Condon factors, but limits our study to weak laser fields only,
where only a single photon is exchanged with the field.  Typical
cooling lasers are strong and detuned only a few linewidths.  By
alternating with a weaker probe laser one can access the particular
range of detuning and intensity that we study.  Future studies are 
needed to address the effects of strong laser fields and the 
consequent revision of our results due to saturation and power 
broadening~\cite{Holland94,Suominen98}.

We also include rotational states into our model, which for the
molecular ground state correspond to the partial waves of a standard
scattering problem (angular momentum quantum number $l$).  The
symmetry of spinless alkaline earth dimers permits only even partial
waves.  It should be pointed out that for near-resonant light the
Condon point is at relatively large distances.  This means that
although the collision energy is low, one needs to go to relatively
large $l$ before the ground state centrifugal barrier stops the
quasimolecule from reaching $R_C$.  We correct here some mistakes that we
discovered in the sum over partial waves in Ref.~\cite{Machholm99}.

The probe laser can be tuned over a wide range from a few to many
atomic linewidths.  For sufficiently large detuning the rotational
structure becomes sharp, and it should be possible to resolve the
vibrational and rotational states.  However, even at 27 GHz detuning,
rotational features in the Ca$_2$ ${^1}\Sigma_g\to{^1}\Sigma_u$
photoassociation spectrum are only partially
resolved~\cite{Tiemann00}.  Photoassociation studies can yield precise
information on the molecular potentials.  Also, the photoassociation
line shapes are sensitive to the near-threshold ground state wave
function, especially if it has nodes in the region swept by the
detuning-dependent Condon point.  Analysis of such data would
hopefully give a value for the $s$-wave scattering length for Mg or
other alkaline earth species, and consequently determine whether a
stable Bose-Einstein condensate is possible.

In this paper we present the calculated estimates for trap loss rate
coefficients in Mg at temperatures around and below the Doppler
temperature.  The contributions from different mechanisms and states
are identified and compared.  We also calculate predictions for other
alkaline earth atoms and Yb by combining the appropriate atomic
properties with model molecular potentials.  Sec.~\ref{MolPhys}
presents in detail the atomic data, molecular potentials, and laser
couplings used for Mg and other alkaline earth atoms. 
Sec.~\ref{methods} describes our theoretical approach.  The results
for Mg are given in Sec.~\ref{resultsMg} and for the other atoms (Sr,
Ca, Ba, Yb) in Sec.~\ref{resultsOther}.  Finally we present some
conclusions in Sec.~\ref{conclusions}.

\section{The molecular physics of alkaline earth dimer molecules}
\label{MolPhys}

\subsection{Alkaline earth atomic structure}

\begin{figure}[htb]
%\vspace{7cm}
\noindent\centerline{
\psfig{width=85mm,figure=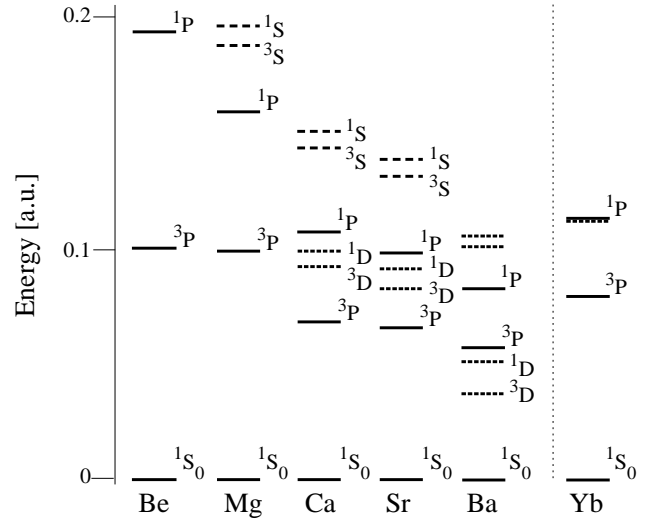}
}
\caption[f2]{Energy levels of Group II atoms and Yb.
\label{atomiclevels}}
\end{figure}

Table~\ref{tbl1} gives the basic atomic data for the alkaline earth
atoms and Yb, which has an electronic structure similar to the Group
II elements.  The major isotopes have no hyperfine structure,
except for Be.  Alkaline-earth atoms have a $^{1}$S$_0$ ground state,
and excited $^{1}$P$_1$ and $^{3}$P$_1$ states that are optically
connected to the ground state, the latter weakly.  Figure
\ref{atomiclevels} sketches the energy levels of the Group II atoms.
The $^{1,3}$P first excited states are the most important for laser
cooling.  The $^{1,3}$D states shift downwards as the atomic number
increases.  For Mg the $^{1,3}$D states are above the $^{1,3}$P
states.  For Ca and Sr they are between the $^{3}$P and the $^{1}$P
states, and for Ba the $^{1,3}$D states are below the $^{1,3}$P
states.  In laser cooling one uses typically the $^1$S$_0$-$^1$P$_1$
transition, which is the situation studied in this paper.  For Mg this
requires a UV laser source, and for the heavier elements requires
repumping to recycle atoms that decay to lower levels.  The weak
$^1$S$_0$-$^3$P$_1$ intercombination transition has a very narrow
linewidth, and is within the optical range.  Thus alkaline earth atoms
are good candidates for optical atomic clocks when cooled to low
temperatures.  For clock applications we need to understand their
laser cooling properties, including the magnitude and nature of
laser-induced collisional trap loss.

Beryllium is not likely to be a serious candidate for laser cooling. 
It is toxic, has a very short wavelength cooling transition, and the
intercombination ``clock'' transition is so weak as to be effectively
forbidden.  Therefore, we will not consider Be in the rest of this
paper.

\newpage
\widetext

%Table I goes here
\begin{table}[htb]
\caption[t1]{Atomic data for major isotopes of Group II elements and
Yb without hyperfine structure (natural abundance shown).  Most data
is derived from \cite{atomdata}.  The lifetime $\tau$ for $^{1}$P$_1$
is taken to be the inverse of the $^{1}$P$_1$-$^1$S$_0$ spontaneous
emission rate $\Gamma_{\mrm at}/\hbar$, thus neglecting weak
transitions to other states for Ba and Yb.  The $^{3}$P lifetimes are
from several sources \cite{datatriplets}.  The linewidth in frequency
units is $\Gamma_{\mrm at}/h=(2\pi \tau)^{-1}$.  The wavelengths
$\lambda$ and fine structure splittings for Sr and Ba are taken from
\cite{Radzig}.  The Doppler temperature is defined as
$T_D=\Gamma_{at}/(2k_B)$.  We take the recoil temperature to be
$T_R=(\hbar^2)/(m \lambdabar^2 k_B)$, where $\lambdabar=\lambda/2\pi$
and $m$ is the atomic mass.  The dipole moment is $d_0=\sqrt{3
\Gamma_{\mrm at}\lambdabar^3 /4}$ (in a.u.).  The atomic units for
dipole moment, length, and energy are $e a_0 = 8.4783 \times 10^{-30}$
Cm, $a_0 = 0.0529177$ nm, and $e^2/(4 \pi \epsilon_0 a_0) = 4.3597482
\times 10^{-18} $ J respectively.
\label{tbl1}}
\begin{tabular}{l|rrrrrr}
& Be & Mg & Ca & Sr & Ba & Yb\\
\hline
Major isotopes without\\
hyperfine
structure
&${^9}$Be(100\%)&$^{24}$Mg(78.99\%)&$^{40}$Ca(96.94\%)&$^{88}$Sr(82.58\%)
       &$^{138}$Ba(71.70\%)&$^{174}$Yb(31.8\%)\\
(abundance)&&$^{26}$Mg(11.01\%)&&$^{86}$Sr(9.86\%)&&$^{172}$Yb(21.9\%)\\
\hline
$\tau$\\
$^{1}$P$_1$ (ns) & 1.80 & 2.02 & 4.59 & 4.98 & 8.40 & 5.68\\
$^{3}$P$_1$ (ms) &  & 5.1 & 0.48 & 0.021 & 0.0014&0.00088\\
\hline
$\Gamma_{\mrm at}/h$\\
$^{1}$P$_1$ (MHz) & 88.5 & 78.8 & 34.7 & 32.0 & 18.9 & 28.0\\
$^{3}$P$_1$ (kHz) & & 0.031 & 0.33 & 7.5 & 120& 181\\
\hline
Doppler-cooling limit\\
$^{1}$P$_1$ (mK) & 2.1 & 1.9 & 0.83 & 0.77 & 0.45 & 0.67\\
$^{3}$P$_1$ (nK) & & 0.75 & 8.0 & 179 & 2.8 10$^{3}$ & 4.4 10$^{3}$\\
\hline
Recoil limit\\
$^{1}$P$_1$ ($\mu$K) & 39 & 9.8 & 2.7 & 1.0 & 0.45 & 0.69\\
$^{3}$P$_1$ ($\mu$K) & & 3.8 & 1.1 & 0.46 & 0.22 & 0.36\\
\hline
$d_0$ (a.u.)\\
$^{1}$S$_0-^{1}$P$_1$ & 1.89 & 2.38 & 2.85 & 3.11 & 3.16 & 2.35\\
\hline
$\lambda$ (nm)\\
$^{1}$S$_0-^{1}$P$_1$ & 234.861 & 285.21261 & 422.6728 & 460.733 &
553.548 & 398.799\\
\hline
$\lambdabar=\lambda/(2\pi)$ (a$_0$)\\
$^{1}$S$_0-^{1}$P$_1$ & 706.4 & 857.8 & 1271.2 & 1385.7 & 1664.9 &
1199.4\\
\hline
FS splitting\\
$^3$P$_2-^3$P$_0$ (a.u.) &
1.36 10$^{-5}$ & 2.77 10$^{-4}$ &
7.20 10$^{-4}$ & 2.65 10$^{-3}$ & 5.69 10$^{-3}$ & 1.10 10$^{-2}$
\end{tabular}
\end{table}

\narrowtext

Because of the lack of hyperfine structure the basic laser cooling
mechanism for alkaline earth atoms is Doppler cooling, for which the
temperature limit $T_D$, defined in the caption of Table ~\ref{tbl1},
is set by the linewidth $\Gamma_{\mrm at}$ of the cooling transition
(widths in this paper are expressed in energy units, so that the decay
rate is $\Gamma_{\mrm at}/\hbar$).  The lifetime of the alkaline earth
$^{1}$P$_1$ state is between 1.8 and 8.4 ns, giving a Doppler-cooling
limit between 2.1 and 0.45 mK for the elements in Table~\ref{tbl1}. 
On the other hand the $^{3}$P$_1$ state has a long lifetime with
Doppler-cooling limits in the nK range.  This can be compared to the
photon recoil limit $T_R$, defined in the Table caption; Table
\ref{tbl1} shows that $T_R$ is between 0.2 and 4 $\mu$K. Thus the
recoil limit is above the intercombination line Doppler cooling limit
for Mg and Ca, nearly coincident with it for Sr, and below it for Ba
and Yb.

Although Sisyphus cooling and magnetic trapping is not available for
${^1}$S$_0$ atoms as it is for alkali atoms, intercombination line
cooling is possible for some Group II species.  This has been used to
cool Sr to $\sim 400$ nK with relatively high phase space
density $> 0.1$ \cite{Katori99,Katori99b}.  If combined with far
off-resonant optical traps and evaporative cooling it may become
possible to obtain Bose-Einstein condensation with optical methods
alone.

\newpage
\vspace*{14cm}

\subsection{Alkaline earth dimer molecular structure}\label{structure}

\begin{figure}[htb]
%\vspace{7cm}
\noindent\centerline{ \psfig{width=85mm,figure=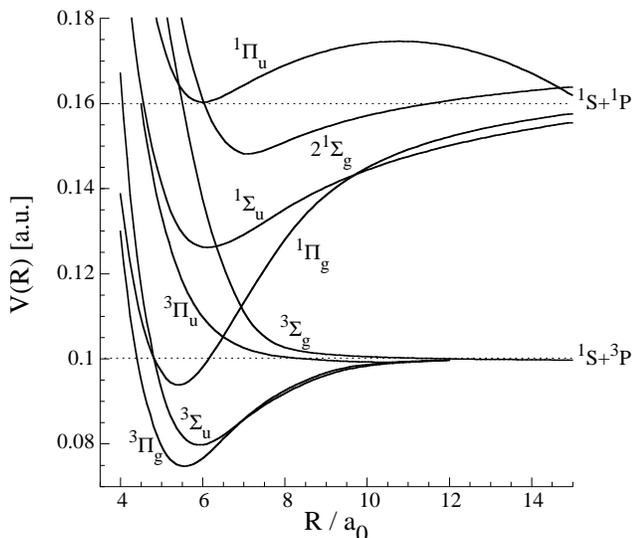} }
\caption[f3]{The molecular states of Mg$_2$ in atomic units
corresponding to the asymptotic atomic states $^1$S$_0$+$^1$P$_1$ and
$^1$S$_0$+$^3$P$_{0,1,2}$~\cite{Stevens77}; the zero of energy is at
the ground state $^1$S$_0$+$^1$S$_0$ asymptote.  There are four states
correlating with each asymptote, of which two are attractive and two
are repulsive at large $R$, where the system is expected to be
resonant with the laser field.
\label{potentials}}
\end{figure}

Figure~\ref{potentials} shows the lowest electronic potentials for
Mg$_2$~\cite{Stevens77}.  There are only two states with attractive
long-range potentials correlating with ${^1}$P$_1+{^1}$S$_0$
that can be resonantly coupled to the ground state by laser
light, namely
$^1\Sigma^+_u$ and $^1\Pi_g$.  Both states offer the possibility for
the SC and RE mechanisms.  When comparing Figs.~\ref{atomiclevels}
and~\ref{potentials} one can see that Mg is special.  For other
alkaline earth atoms the molecular state picture is further
complicated by the atomic D states below the ${^1}$P$_1$ state.  This
increases the number of molecular states and thus the number of
energetically available exit channels.  The small number of molecular
states is one reason we have chosen Mg as the basis for our studies.
Theoretical calculations require precise information about the
molecular potentials and couplings over a wide range of interatomic
distance.  {\it Ab initio} calculations of ground and excited
molecular potentials are also available for Sr~\cite{Boutassetta96}
and Ba~\cite{Allouche95}.

While working on the manuscript we received new {\it ab initio} data
on Mg$_2$ from E. Czuchaj, including both improved potential curves
and spin-orbit couplings~\cite{Czuchaj}.  Although the new data differ
to some extent from the results of Stevens and Krauss~\cite{Stevens77}
because of improved calculations of electron correlation effects, we
do not expect that use of the new data would lead to any strong
modification of our basic results, which should be viewed as
qualitative model calculations for the reasons to be discussed in the
following sections.

The linewidth of the excited ${^1}\Sigma_u$ state depends on the
interatomic distance $R$ with a magnitude on the order of twice the
atomic linewidth.  Thus the vibrational levels of the ${^1}\Sigma_u$
state near the ${^1}$S$+{^1}$P dissociation limit overlap strongly,
and one does not expect to resolve them.  One interesting point is
that the dipole-forbidden ${^1}\Pi_g-{^1}\Sigma_g$ transition becomes
allowed at large $R$ due to retardation corrections.  This means that
the ${^1}\Pi_g$ state can be excited at $R_C$, but the spontaneous
emission probability goes down quickly as $R$ decreases.
Consequently, ${^1}\Pi_g$ vibrational states near the dissociation
limit have narrow emission linewidths.  Thus the assumption that the
vibrational states overlap strongly and can not be resolved at
detunings of a few linewidths is not necessarily valid.  One must
determine if resolvable features persist when one sums over all
involved rotational states/partial waves, and takes the energy average
over a thermal distribution.  We show that one can indeed see
vibrational structure, especially if the temperature is well below the
$^1$S$_0$-$^1$P$_1$ Doppler limit.  As mentioned above, this is by no
means impossible, if one does the cooling using the
$^1$S$_0$-$^3$P$_1$ transition.

\subsection{Long-range properties of states correlating with
$^{1}$P$_1+^{1}$S$_0$ atoms}\label{LongRange}

There are four molecular states correlating with
$^{1}$P$_1$+$^{1}$S$_0$ atoms: two long-range attractive states,
$^1\Sigma^+_u$ and $^1\Pi_g$, and two long-range repulsive states,
$^1\Sigma^+_g$ and $^1\Pi_u$.  Here long-range means that $R$ is large
compared to the short-range region of chemical bonding and van der
Waals interactions shown in Fig.~\ref{potentials} so that the
potential is determined by the first-order dipole-dipole-interaction
with $C_3 = \mp 2d_0^2$ and $\mp d_0^2$ for the $^1\Sigma$ and $^1\Pi$
states, respectively.  Here $d_0$ is the $z$-component of the atomic
transition dipole matrix element, which is related to the atomic
linewidth $\Gamma_{\mrm at}$ and the reduced wavelength of the
$^{1}$S$_0-^{1}$P$_1$ transition ($\lambdabar=\lambda/2\pi$) by
$d_0^2=3\lambdabar^3\Gamma_{\mrm at}/4$.  The exact long-range (lr)
potentials including relativistic retardation corrections
are~\cite{Meath68}
\begin{eqnarray}
   \label{longrangepot}
   V_{\mrm lr}(u;{^1\Sigma^+_u})&=&-\frac{3\Gamma_{\mrm at}}{2u^3}
   [\cos(u)+u\sin(u)],\\
   V_{{\mrm lr}}(u;{^1\Pi_g})&=&-\frac{3\Gamma_{\mrm at}}{4u^3}
   [\cos(u)+u\sin(u)-u^2\cos(u)] , \nonumber
\end{eqnarray}
where $u=R/\lambdabar$ is the scaled distance.  The
molecular linewidths with relativistic retardation corrections
are~\cite{Meath68}
\begin{eqnarray}
   \label{mollinewidth}
   \Gamma(u;{^1\Sigma^+_u})&=&
        \Gamma_{\mrm at}\left\{1-\frac{3}{u^3}
          \left[u\cos(u)-\sin(u)\right]\right\},\\
   \Gamma(u;{^1\Pi_g})&=&
        \Gamma_{\mrm at}\left\{1-\frac{3}{2u^3}
          \left[u\cos(u)-(1-u^2)\sin(u)\right]\right\}.\nonumber
\end{eqnarray}
In the region with $R < \lambdabar$, the potentials vary as $1/u^3$,
and $\Gamma({^1\Sigma^+_u})$ and $\Gamma({^1\Pi_g})$ respectively vary
as $2\Gamma_{\mrm at}$ and $\Gamma_{\mrm at}u^2/5$.

The very long-range excited state potentials result in large Condon
points for excitation of the attractive states.  If the laser detuning
relative to the atomic transition is expressed in units of
$\Gamma_{\mrm at}$ and the distance in units of $\lambdabar$, the
scaled Condon point ($u_C$) for the ground-excited state transition
becomes independent of atomic species.  Table~\ref{Condonpoints} shows
$u_C$ for several detunings $\Delta$, where we define red detuning to
be positive.  For Mg at $\Delta$=$\Gamma_{\mrm at}$ we have
$R_{C}(^1\Sigma^+_u)$=1132 a$_0$ and $R_{C}(^1\Pi_g)$=728
a$_0$.

\begin{table}[htb]
\caption[t1]{Condon points in scaled distance for selected detunings
\label{Condonpoints}}
\begin{tabular}{rdd}
Detuning $\Delta$ & $u_{C}({^1\Sigma^+_u})$ & $u_{C}({^1\Pi_g})$\\
\hline
$\Gamma_{\mrm at}$    & 1.32  & 0.849 \\
5$\Gamma_{\mrm at}$   & 0.716 & 0.511 \\
10$\Gamma_{\mrm at}$  & 0.555 & 0.411 \\
30$\Gamma_{\mrm at}$  & 0.376 & 0.288 \\
\end{tabular}
\end{table}

The attractive potentials support a series of vibrational levels
leading up to the dissociation limit.  Assuming a potential with a
long-range form  $-C_3/R^3$ gives the binding energy of
vibrational level $v$~\cite{LeroyBernstein}:
\begin{equation}
     \varepsilon_v^{1/3} = \left ( \frac{\pi}{2a_3}\right)^2
     \frac{\hbar^2}{2\mu C_3^{2/3}} (v -v_D)^2,
     \label{LB1}
\end{equation}
where $a_3 = (\sqrt{\pi}/2)\Gamma(5/6)/\Gamma(4/3)
= 1.120251$,
$v_D$ is the vibrational quantum number at the dissociation limit
($v_D$ is generally nonintegral) and $\mu$ is the reduced mass.  The
vibrational spacing function, which we need later, is
\begin{equation}
        \frac{\partial\varepsilon_v}{\partial v} = h \nu_v =
        \frac{3\pi}{a_3} \left(\frac{\hbar^2}{2\mu C_3^{2/3}}\right)^{1/2}
        \varepsilon_v^{5/6} ,
        \label{LB2}
\end{equation}
where $h\nu_v$ is the vibrational frequency.

\subsection{Ground state}\label{GroundState}

The ground-state van der Waals potential varies at long-range as
$R^{-6}$ and is essentially flat for the range of Condon points we
consider.  The energy-normalized ground state scattering wave function
for collisional momentum $\hbar k_\infty$ and partial wave $\l$ has
the long-range form
\begin{equation}
	  \Psi(R,\l,k_\infty) = \left (\frac{2\mu}{\pi \hbar^2 k_\infty}
  	\right)^\frac{1}{2} \sin\left[ k_\infty R-\frac{\pi}{2}\l
	  +\eta_\l(k_\infty)\right ] .
   \label{gspsi}
\end{equation}
We define the collisional energy as $\varepsilon=\hbar^2k^2_{\infty}/(2\mu)$.
The short-range potentials are not sufficiently well-known for
alkaline earth dimers to determine accurately the scattering phase
shifts $\eta_\l(k_\infty)$.  Therefore, our calculations will have to
be model calculations.  However, we test the sensitivity of our
trap loss spectra to the unknown phases, and show that this is not a
serious limitation.  There are two reasons for this.  One is that the
ground state potential is flat in the long-range region, and the
amplitude of $\Psi(R,\l,k_\infty)$ has its asymptotic value in
Eq.~(\ref{gspsi}) independent of $R$ as long as $R > x_0 =
\frac{1}{2}(2 \mu C_6/\hbar^2)^{1/4}$~\cite{Julienne96,Boisseau00};
the use of $x_0$ (or a closely related length) as an appropriate
length scale for van der Waals potentials is described in the Appendix
of Ref.~\cite{Williams99}.  The condition $R_C \gg x_0$ is easily
satisfied in our case.  Second, at the Doppler limit $T_D$ for
${^1}$S$_0 \to {^1}$P$_1$ cooling, a number of partial waves $\l$
contribute to trap loss in our detuning range (1-50 $\Gamma_{\mrm
at}$).  We demonstrate in Sec.~\ref{MgT_D} that a sum over $\l$
removes the dependence on the short-range potential.

Our trap loss spectra for $s$-wave scattering at the low temperatures
available via intercombination line cooling will be sensitive to the
actual scattering length $A_0$ of the ground ${^1}\Sigma_g^+$
potential.  However, we demonstrate a simple scaling relationship that
will allow our low $T$ $s$-wave results to be scaled to any value of
the scattering length.  We may expect that the ${^1}\Sigma_g^+$
scattering length can be determined from one or two-color
photoassociation spectra, as has been done for alkali
species~\cite{Weiner99}.  However, such analysis will require an
accurate $C_6$ coefficient and for optimum results needs a reasonably
accurate short-range potential as well.

\subsection{Excited state short-range potentials}
\label{ShortRange}

Trap loss spectra depend strongly on the excited state short-range
potential structure in three ways:

(1) The curve crossings leading to SC trap loss occur at short range,
and determine $P_{pe}$.  In a Landau-Zener interpretation,
\begin{equation}
     P_{pe}(J') =
     4e^{-2\pi\Lambda}\left(1-e^{-2\pi\Lambda}\right)
     \sin^2(\beta_{J'}) ,
     \label{LZ}
\end{equation}
where $\Lambda = |V_{pe}(R_p)|^2 /(\hbar v_p D_p)$ and $J'$ designates
excited state total angular momentum (See Section \ref{MolRot}).  Here
$V_{pe}(R_p)$, $v_p$, $D_p$ are the respective coupling matrix
element, speed, and slope difference at the $R_p$ crossing, and
$\beta_{J'}$ is a semiclassical phase angle~\cite{Nikitin}:
\begin{equation}
        \beta_{J'}=\int_{R_{0e}}^{R_p} k_e(R,{J'}) dR - \int_{R_{0p}}^{R_p}
        k_p(R,{J'}) dR+ \frac{\pi}{4} ,
        \label{LZ2}
\end{equation}
where $R_{0i}$ and $\hbar k_i(R,{J'})$ are the respective inner classical
turning point at zero energy and local momentum for state $i=$ $e$ or
$p$.  The attractive singlet potentials may have one or more crossings
with repulsive potentials from lower-lying states, e.g., those
correlating to $^3$P$+^1$S. Thus the number, positions of crossings
and the coupling between states with crossing potentials are important
for the magnitude of $P_{pe}$.  Although we can make reasonable
estimates for $\Lambda$, the phase angle $\beta_{J'}$ is sensitive to
details of the potentials and can only be calculated accurately if
very accurate potentials are available~\cite{Dulieu94}.

(2) The vibrational structure in the trap loss spectra depends on both
short and long-range potentials.  The spacing between the vibrational
levels given by Eq.~(\ref{LB2}) depends only on the long-range
potentials, but the exact positions of the levels in Eq.~(\ref{LB1}) are
determined by the short-range potentials in the region of chemical
bonding (through the $v_D$ parameter).  Thus, the magnitude of the
vibrational spacings in our model calculations will be correct,
whereas the actual positions can only be determined by measurement.

(3) The short-range SC process also contributes to the width of
vibrational features in the trap loss spectra \cite{Napolitano94}.
Depending on species, temperature, and detuning, the widths may be
primarily determined by natural or thermal broadening or by the
predissociation decay rate related to $P_{pe}$.  However, we are able
to place approximate bounds on the magnitude of $P_{pe}$.
Sections~\ref{LimitingCases}, \ref{resultsMg}, and \ref{resultsOther}
discuss the contributions to feature widths and show that natural and
thermal broadening tends to be dominant at small detuning, whereas SC
broadening may become dominant at large detuning.

The available data on short-range potentials varies among the Group II
elements.  {\it Ab initio} potentials are available for
Mg~\cite{Stevens77,Czuchaj}.  The structure of the molecular
potentials is fairly simple because only states correlating to
$^1$P+$^1$S (4 states) and $^3$P+$^1$S atoms (4 states) are present
(see Fig.  \ref{potentials}) and the potentials provide qualitative
data for possible SC mechanisms.  On the other hand, Ca, Sr and Ba
have a very complex short-range structure, because of large fine
structure splittings in the triplet states, and the states correlating
to $^1$D+$^1$S and $^3$D+$^1$S coming into play.  For example,
Boutassetta {\it et al.}~\cite{Boutassetta96} and Allouche {\it et
al.}~\cite{Allouche95} calculated the short-range potentials for Sr
and Ba respectively, but the large number of states (e.g., 38
states for Ba) makes it excessively complicated to treat the
short-range SC mechanism even qualitatively.

The coupling between the singlet and triplet states at short-range is
due to spin-orbit couplings.  The exact magnitude of the couplings is
unknown for all Group II elements, but can be estimated using Table A1
of Ref.~\cite{Hay76}, which relates the coupling matrix elements to
the $^3$P$_2-^3$P$_0$ fine structure splitting.  This approximation
ignores any $R$-dependence of the spin-orbit matrix elements due to
chemical bonding.  The fine structure splitting of the $^3$P state
increases with atomic number (see Table \ref{tbl1}).

We have opted for Mg as a model system, because short-range potentials
are available.  Each singlet state with attractive potential couples
to only one triplet state in the inner zone.  Thus the SC trap loss
problem for Mg decouples into two three-state calculations, one for
the $^1\Sigma^+_u$ and one for the $^1\Pi_g$ excited state.  For Mg we
provide qualitatively correct SC trap loss spectra and even give
some quantitative estimates.

In the case of the other Group II elements and Yb, we treat their
complicated inner zone physics as one effective crossing, that is, we
use three-state calculations based on the Mg-model, and explore the
effects of mass, radiative properties, and coupling strength (size of
spin-orbit splitting) in these model calculations.  Thus we do not
use the {\it ab initio} potentials of
Refs.~\cite{Boutassetta96,Allouche95} discussed in the paragraph
above, because even if we were to include all the curves, there would
still be uncertainties associated with unknown phases and unknown
coupling matrix elements.  Rather, our goal is to indicate qualitative
differences in magnitude and spectral shapes among the various
species.  We trust that these will be helpful in providing guidance
for future experimental and theoretical studies of these systems.  As
we discuss in Section \ref{Factorization}, our calculation of the
excitation-transfer coefficient $\kappa$ will be to a large extent
independent of details of the complicated short-range molecular
physics and curve crossings.

\subsection{Molecular rotational structure and coupling to the laser field}
\label{MolRot}

The full three-dimensional treatment of the collision of a $^1$S atom
coupled to a $^1$P atom by a light field is worked out in
Ref.~\cite{Napolitano96}.  We adapt this treatment to our simplified
model with molecular Hund's case (a) transitions between two molecular
states with the usual rotational branch structure.  The angular
momentum $J''$ in the ground state can only be that of molecular axis
rotation, $J''=\l''$ with projection $m''$ on a space-fixed axis.  The
excited rotational levels in a Hund's case (a) molecular basis for a
$^1\Sigma^+_u$ or $^1\Pi_g$ state do not have mechanical rotation
$\l'$ as a good quantum number but instead the total angular momentum $J'$
with space projection $M'$.  The three possible transition branches
have $J'=\l''+B$, where the branch labels P, Q, and R
respectively designate the cases $B =$ -1, 0, and 1.  The
quasi-molecule ground state can couple to the $^1\Sigma^+_u$ state
only by P and R branches, but to the $^1\Pi_g$ state by P, Q,
and R branches.  All potentials have a centrifugal term added,
$n(n+1)/(2\mu R^2)$, depending on ground state $n=\l''$ or
excited states $n=J'=\l''+B$.

The radiative coupling terms are
\begin{equation}
        V_{A,B,l'',m'',q}(u) = \left ( \frac{2 \pi I}{c} \right)^{1/2}
           \langle A,J'M'| \hat{e}_q \cdot \vec{d} | \l''m''\rangle ,
           \label{Rad}
\end{equation}
where $A =$ $^1\Sigma_u^+$ or $^1\Pi_g$ labels the excited state, $I$
is light intensity, $\vec{d}$ is the lab frame dipole operator,
$\hat{e}_q$ is the polarization vector of light with polarization $q =
0,\pm 1$, and $M'$, $m''$ are lab frame angular momentum projection
quantum numbers.  In our weak-field case, where transition
probabilities are linear in $|V_{A,B,l'',m''(u),q}|^2$, we can define
an $m''$-averaged radiative coupling matrix element (the sum over $m''$
removes the dependence on $q$):
\begin{eqnarray}
   &&V_{eg,A}(u,\l'',B,I)\nonumber\\
   &&= \left ( \frac{1}{2\l''+1} \sum_{m''=-\l''}^{l''}
   |V_{A,B,l'',m'',q}(u)|^2 \right )^{1/2} \nonumber \\
   &&=\left ( \frac{2 \pi I}{c} \right)^{1/2}
   \alpha_{A,B,l''} d_A(u)\nonumber \\
   &&=2.669\times10^{-9}\alpha_{A,B,\l''} \sqrt{I({\mrm W/cm}^2)} d_A(u)
   ({\mrm a.u.}) .\nonumber\\
\end{eqnarray}
The molecular electronic transition dipole moment $d_A$ is
\begin{equation}
        d_A(u) ({\mrm a.u.}) = \sqrt{\frac{3\lambdabar({\mrm
        a.u.})^3}{4}\Gamma_A(u) ({\mrm a.u.})}.
        \nonumber
\end{equation}
where $\Gamma_A(u)$ is the molecular linewidth of the excited state as
in Eq.~(\ref{mollinewidth}).  The factors $\sqrt{2\l''+1} \,\alpha_{A,B,\l''}$
are given in Table~\ref{alphafactor}.

%Table III goes here
\begin{table}[htb]
\caption[t1]{The rotational line strength factors
$\sqrt{2\l''+1}\,\,\alpha_{A,B,l''}$.
\label{alphafactor}}
\begin{tabular}{dddd}
State $A$ & Branch $B$ & $l''=0$ ($s$-wave) & $l''\neq 0$\\
\hline
$\Sigma$ & P & 0            & $\sqrt{l''/3}$      \\
$\Sigma$ & R & $\sqrt{2/3}$ & $\sqrt{(l''+1)/3}$  \\
$\Pi$    & P & 0            & $\sqrt{(l''-1)/3}$  \\
$\Pi$    & Q & 0            & $\sqrt{(2l''+1)/3}$ \\
$\Pi$    & R & $2/\sqrt{3}$ & $\sqrt{(l''+2)/3}$  \\
\end{tabular}
\end{table}

\subsection{Model potentials for Mg}\label{MgModel}

Our results are not sensitive to the detailed form of the ground state
potential, for the reasons given in Sec.~\ref{GroundState}.  Thus,
we model the ground state potential by a Lennard-Jones 6-12 form
\begin{equation}
   \label{shortrangepot}
   V_g(u)=4 \epsilon \left[ \left (\frac{\sigma}{\lambdabar
   u}\right)^{12}-\left(\frac{\sigma}{\lambdabar u}\right)^6
   \right] .
   \label{LJ6-12}
\end{equation}
For Mg we model the ground state potential from \cite{Stevens77} with
a well depth of $\epsilon=0.002825$ a.u.\ and an inner turning point
of $\sigma=6.23$ a$_0$.  This potential has a scattering length of
$-95$ a$_0$ for the $^{24}$Mg reduced mass of 23.985042/2 atomic mass
units.  The scattering length $A_0$ is determined from the $k \to 0$
behavior of the $s$-wave phase shift: $\eta_0 = -k_\infty A_0$.

Since the excited {\it ab initio} potentials of Ref.~\cite{Stevens77}
do not permit a quantitative calculation of spectroscopic accuracy,
for the reasons given in Sec.~\ref{ShortRange}, the specific forms of
the short-range (sr) excited-state potentials are not important for
our purposes of modeling the qualitative structure and magnitude of
the collisional loss.  However, it is important to retain the correct
long-range form.  Therefore for simplicity in the calculations and
because we want to model the other alkaline earth systems to explore
the effect of different $C_3$, $\Gamma_{\mrm at}$, $\lambdabar$ and
mass, we have modeled the {\it ab initio} potentials with
Lennard-Jones 3-6 potentials keeping the long-range form fixed to its
known $C_3$ value given in Eq.~(\ref{longrangepot})
\begin{equation}
   \label{shortrangepot2}
   V_{\mrm sr}(u)=4 \epsilon \left[ \left
   (\frac{\sigma}{\lambdabar u}\right)^6-\left(\frac{\sigma}{\lambdabar
   u}\right)^3 \right].
\end{equation}

We have two fitting parameters $\epsilon$ and $\sigma$, and three
given values: well depth of the {\it ab initio} potential (the depth of
the model potential is $\epsilon$), position of minimum $u_{\mrm min}$
and $C_3({^1}\Sigma^+_u)=-2 d_0^2$ or $C_3({^1}\Pi_g)=- d_0^2$.  Because
we want to fix the long-range potential $C_3$ we can not fit $u_{\mrm
min}$ and the well depth at the same time and have chosen the latter:
\begin{eqnarray}
   \label{excitedpot1}
   V_e(u;{^1\Sigma^+_u})&=&4\epsilon(\Sigma)
   \left(\frac{\sigma(\Sigma)}{\lambdabar u}\right)^6\\
   &&-\frac{2d^2_0}{\lambdabar^3u^3}[\cos(u)+u\sin(u)] ,\nonumber\\
   V_e(u;{^1\Pi_g})&=&4\epsilon(\Pi)
   \left(\frac{\sigma(\Pi)}{\lambdabar u}\right)^6\label{excitedpot2}\\
   &&-\frac{d^2_0}{\lambdabar^3u^3}[\cos(u)+u\sin(u)-u^2\cos(u)],\nonumber
\end{eqnarray}
where

\begin{tabular}{lll}
$\epsilon(\Sigma)=0.0347$ a.u.,& &$\sigma(\Sigma)=4.339$ a$_0$,\\
$\epsilon(\Pi)=0.0681$ a.u., & &$\sigma(\Pi)=2.751$ a$_0$.\\
\end{tabular}

\noindent The well minima are $\lambdabar u_{\mrm min}=5.5$ a$_0$ and
3.5 a$_0$ for $^1\Sigma^+_u$ and $^1\Pi_g$, respectively, compared to
the {\it ab initio} values $\lambdabar u_{\mrm min}=6.1$ a$_0$ and 5.4
a$_0$.

Both excited states have a SC mechanism in the short-range region with
coupling to a triplet state. The triplet states are purely repulsive, and
modeled with
\begin{equation}
   V_{p,A}(u)=\frac{C_6}{\lambdabar^6 u^6}+V_{\infty} ,
\end{equation}
where $A=$ ${^1}\Sigma_u$ or ${^1}\Pi_g$ labels the molecular state,
and $V_{\infty}=-0.0601$ a.u.

The SC from $^1\Sigma^+_u$ to $^3\Pi_u$ takes place around the inner
turning point of the $^1\Sigma^+_u$ potential well.  We have chosen
$C_6=392$ a.u. for the model of the $^3\Pi_u$ state.

The SC from $^1\Pi_g$ to $^3\Sigma_g$ takes place about 1.5 a$_0$
outside and 0.019 a.u.\ above the minimum of the $^1\Pi_g$ state
potential well.  With $C_6=81$ a.u.\ we have a model of the crossing
where the corresponding values are: 1.0 a$_0$ and 0.019 a.u. The
difference in slope of the crossing potentials is 0.030 a.u.\ for the
{\it ab initio} potentials and 0.037 a.u.\ for the model.

The coupling between the crossing states are approximated using Table
A1 of Ref.~\cite{Hay76}.  The $^1\Sigma^+_u$-$^3\Pi_u$ and
$^1\Pi_g$-$^3\Sigma^+_g$ matrix elements are $\zeta/\sqrt{2}$ and
$\zeta/2$, respectively, where $\zeta=1.84\times10^{-4}$ a.u.\ is 2/3
of the atomic $^3$P$_2$-$^3$P$_0$ splitting.  For example, we estimate
an upper bound ($\sin^2\beta_{J'}=1$) to the Landau-Zener version of
$P_{pe}$ in Eq.~(\ref{LZ}) for the $^1\Pi_g$-$^3\Sigma_g$ crossing to be
2.6$\times 10^{-3}$ for the model potentials and 2.8$\times
10^{-3}$ estimated from the {\it ab initio} potentials.
This upper bound is consistent with the calculated $P_{pe}$
as a function of $J'$ from our complex potential calculation
described below.

Since we will use a semiclassical method to determine the $P_{pe}$
factor for the RE process via the $^1\Sigma^+_u$ state (see
Sec.~\ref{REcalculations} below), we do not need an explicit probe
channel for RE. However, we introduce a probe channel to simulate
RE in order to show that the same $J_{eg}$ factor in
Eq.~(\ref{factor1}) applies for both SC and RE processes, irrespective
of the choice of the short-range $R_p$.  Since we may take any
form we like for a RE probe state, we use a probe state
potential which crosses the excited state at a distance $u_p$, where the
kinetic energy gained by the collision pair is 1 K, corresponding to a
trap depth of 0.5 K. The RE probe potential has a repulsive inner
wall
\begin{equation}
   V_{\mrm RE,probe}(u)=\frac{C_{12}}{(\lambdabar u)^{12}}-V_{\mrm
   kin,RE,\infty}  ,
\end{equation}
where $V_{\mrm kin,RE,\infty}=3.17\times 10^{-6}$ a.u. and
$C_{12}=5\times 10^{6}$ a.u. No rotational term is included in this
probe channel.  The same probe state potential is used for all
collision energies, which are small (mK range and below) compared to
the 1 K kinetic energy at $u_p$.  The coupling between the excited
state and the probe state is chosen to be weak: $10^{-9}$ a.u.

\subsection{Model potentials for Ca, Sr, Ba and Yb}

Since the different ground state values of $C_6$ and inner potential
shape make no difference for these model studies, for the reasons
given in Sec.~\ref{MgModel}, we take the same ground state
Lennard-Jones 6-12 potential, Eq.~(\ref{LJ6-12}), as in the Mg case to
model the other alkaline earth ground states, and only change the
reduced mass.  This procedure yields respective model scattering
lengths of 67 a$_0$, -65 a$_0$, -41 a$_0$, and 97 a$_0$ for $^{40}$Ca,
$^{88}$Sr, $^{138}$Ba, and $^{174}$Yb.  Thus, $|A_0| \ll R_C$ in all
model cases.

The long-range of the excited state potentials for Ca, Sr, Ba and Yb
is still exact, using the data from Table \ref{tbl1} with the form in
Eq.~(\ref{longrangepot}).  Due to lack of accurate excited state
short-range molecular potentials and because of their more complicated
structure, we model the trap loss for Ca, Sr, Ba and Yb by scaling the
potentials from the Mg model Eqs.~(\ref{excitedpot1}) and~(\ref{excitedpot2}). 
The well depth $\epsilon$ of the $^1\Sigma^+_u$ and $^1\Pi_g$ potentials are
scaled by the size of the singlet-triplet states splitting compared to that
splitting in Mg, e.g.:
\begin{equation}
   \epsilon_{\mrm Ca}=\epsilon_{\mrm Mg} \frac{E({^1}{\mrm P},{\mrm
   Ca})-E({^3}{\mrm P},{\mrm Ca})}{E({^1}{\mrm P},{\mrm Mg})-E({^3}{\mrm
   P},{\mrm Mg})} .
\end{equation}

The short-range structure is treated as one effective crossing.  The
probe states are qualitatively like those in the Mg model.  The
position (in energy) of the crossing between the $^1\Pi_g$ and probe
state potentials is scaled as above.  The $^1\Sigma^+_u$ and probe
state potentials come very close at the inner wall of the
$^1\Sigma^+_u$ potential around the classical turning point
$V_e(u)=\varepsilon$.

The spin-orbit coupling constant $\zeta$ scales with the spin-orbit
splitting of the $^3$P atomic states.  We use the same definition of
the matrix elements as for Mg.  The Landau-Zener adiabaticity
parameter $2\pi\Lambda$ in Eq.~(\ref{LZ}) for Ba and Yb is larger than
unity, leading to a modified shape of the short-range adiabatic
potentials and very small $P_{pe,{\mrm SC}}$.  Thus for Ba the
$^1\Sigma^+_u$-probe state coupling and for Yb the $^1\Pi_g$ and the
$^1\Sigma^+_u$-probe state couplings have been reduced by about a
factor 5 to obtain values of $P_{pe,{\mrm SC}}$ close to unity, in
order to test the limit of very strong broadening of the vibrational
structure.  We believe this limit is physically more realistic.  The
variety of crossings in these systems might lead to a strong SC
process.

\section{Complex potential close coupled calculations}\label{methods}

\subsection{Description of method}

The weak field approximation assumed in this study allows us to apply
a complex potential method \cite{Julienne94,Boesten93}, since
re-excitation of any decayed quasimolecular population can be ignored.
Furthermore, the weak field only couples each ground-state partial
wave to at most 3 rotational states of the $^1\Sigma_u$ or $^1\Pi_g$
state through the P, Q, or R branches.  However, in the weak field the
excited-state rotational states do not couple further to other
ground-state partial waves.  Therefore we can ignore any partial wave
ladder climbing.  Thus, for each trap loss mechanism we have three
dressed states: a ground state $g$, an excited state $e$, and a probe
state, $p$.  We solve the three-channel, time-independent radial
Schr$\ddot{\mrm o}$dinger equation for ground state collision energy
$\varepsilon$, partial wave $\l=\l''$, for each transition branch $B$
and for a given intensity $I$
\begin{equation}
   \frac{d^2}{dR^2}{\bf \phi}(\varepsilon,R)+\frac{2\mu}{\hbar^2}
   \left[\varepsilon{\bf 1}-{\bf
   V}(R,\l,B,\Delta,I)\right]{\bf \phi}(\varepsilon,R)={\bf 0} ,
\end{equation}
${\bf V}$ is the $3\times3$ potential matrix
\begin{eqnarray}
   &&{\bf V}(u,l,B,\Delta,I)= \label{PotMat}\\
   && \left(\begin{array}{ccc}
   \Delta+V_e(u,l,B)-i\frac{\Gamma(u)}{2} & V_{eg}(u,l,B,I) & V_{pe}(u)
   \label{Vmat}\\
   V_{eg}(u,l,B,I) & V_g(u,l) & 0 \\
   V_{pe}(u) & 0 & V_p(u,l,B) \\
   \end{array}\right) . \nonumber
\end{eqnarray}
The elements of ${\bf V}$ are described in Sec.~\ref{MolPhys} above.  A complex
term $-i\Gamma(u)/2$ is added to the excited-state potential to simulate the
effect of excited-state decay.  The full retarded form of the molecular
linewidth, Eq.~(\ref{mollinewidth}), is used.

Application of standard asymptotic scattering boundary conditions to
the three-component state vector ${\bf \phi}$ gives the $S$-matrix
elements $S_{ij}(\varepsilon,\l,B,\Delta,I)$.  If
$\varepsilon>\Delta$, all three channels are open: $i,j =$ $g$, $p$,
or $e$.  When $\varepsilon < \Delta$, as is normally the case in our
model, channel $e$ is closed, and $S_{ij}$ is only defined for $i,j =$
$g$ or $p$.  We choose the light intensity $I$ low enough that the
results are in the weak field limit where the $P_{pg}=|S_{pg}|^2$
matrix element scales linearly in $I$.  Our results are normalized to
a standard intensity of $I = 1$ mW$/$cm$^2$.

We find that we can make a change in the asymptotic shape of the
artificial probe potential to make the model much more manageable
computationally.  The deep potential well of the excited state and the
large kinetic energy in the probe channel require a small stepsize in
$u$ ($\lambdabar\Delta u\approx 0.005$ a$_0$).  However, with the
large range of $u$ ($\lambdabar u_{\mrm max}\approx$1500-3000 a$_0$) a
small $\Delta u$ increases the computation time and may compromise the
numerical stability.  Therefore, we modify the probe state potential
to bring $V_p(u)$ to a small negative value at intermediate and
asymptotic $u$.  This results in a small asymptotic momentum $\hbar k$
in the probe channel, and allows us to gradually increase the stepsize
to $\lambdabar\Delta u\approx 0.5$ a$_0$ as $u$ increases.  The
coupling between the excited and probe states is turned off
exponentially before the change in $V_p(u)$.  The probabilities
$|S_{pg}|^2$ and $P_{pe}$ are completely independent of the asymptotic
properties of the probe potential if there are no asymptotic barriers.
Since we have no centrifugal potential in the asymptotic probe
channel, there are no asymptotic centrifugal barriers.  The presence
of such barriers in our previously published model~\cite{Machholm99}
resulted in some errors at larger $\l''$ which we have now corrected in
the present model.

The thermally averaged loss rate coefficient via state $e$ is:
\begin{eqnarray}\label{thermalrate}
   K(\Delta,T)=
   &&\frac{k_BT}{hQ_T}\int_0^\infty
     \frac{d\varepsilon}{k_BT}  e^{-\varepsilon/k_BT} \label{KT} \\
   &&\times\sum_{l''_{even},B}(2l''+1)|S_{pg} 
   (\varepsilon,\l'',B,\Delta,I)|^2,
   \nonumber
\end{eqnarray}
where $Q_T=(2\pi\mu k_BT/h^2)^{3/2}$ is the translational
partition function.  Identical particle exchange symmetry ensures that
only even partial waves exist for the ground state.  We also define a
non-averaged rate coefficient for a fixed collision energy
$\varepsilon$, where we define $T_\varepsilon \equiv \varepsilon/k_B$.
Only the sum over partial waves and branches is performed:
\begin{equation}
   \label{sumrate}
   K(\Delta,\varepsilon)=
   \frac{k_B T_\varepsilon}{hQ_{T_\varepsilon}}
   \sum_{l''_{even},B}(2l''+1)|S_{pg}(\varepsilon,\l'',B,\Delta,I)|^2 .
\end{equation}

There are two possible cutoffs $l''_{\mrm max}$ to the partial wave
sum provided by the ground and excited-state centrifugal potentials,
respectively.  For the ground state we can take $l''_{\mrm max}$ to be
the largest integer for which $\hbar^2l''_{\mrm max}(l''_{\mrm
max}+1)/2\mu R_C^2<\varepsilon$ at the Condon point $R_C$.  Thus, the
Condon point is classically accessible for $\l'' \le \l''_{\mrm max}$
and classically forbidden for $\l'' > \l''_{\mrm max}$.  For the
excited state, the centrifugal potential may create a barrier inside
the Condon point for the $g\rightarrow e$ excitation.  The position
and the height of the barrier depend on $J'$.  For collision energies
around $\varepsilon=k_BT_D$ this barrier may prevent allowed
ground-state partial waves from contributing to the loss, because the
excited-state population never reaches the inner zone where RE decay
and SC take place.  In this case, $l''_{\mrm max}$ may decrease from
the value defined by the above inequality.  We find that the ground
state cutoff applies except for the case of high energy and small
detuning.  In either case $|S_{pg}|^2$ decreases many orders of
magnitude as $l''$ varies from $l''_{\mrm max}$ over the next few
$l''$-values.  The upper limit for the sum in Eqs.~(\ref{thermalrate})
and (\ref{sumrate}) is set to the $l''$-value where
$(2\l''+1)|S_{pg}(\varepsilon,\l'',B,\Delta,I)|^2$ is $10^{-6}$ of the
maximum previous $(2\l''+1)|S_{pg}(\varepsilon,\l'',B,\Delta,I)|^2$
value.

\subsection{Factorization of trap loss probability}
\label{Factorization}

The factorization in Eq.~(\ref{factor1}) allows us to separate the
physics of the long-range excitation and the short-range decay to the
trap loss channel.  Reference~\cite{Julienne94} shows how to determine
the short-range probability $P_{pe}$ from a different coupled channels
calculation where the complex decay term $-i \Gamma/2$ in
Eq.~(\ref{PotMat}) is omitted.  Since $P_{pe}$ is determined
in a region near $R_p$ where the local kinetic energy is very high in
relation to $\varepsilon$, this probability is nearly independent of
$\varepsilon$ over a wide range.  Therefore, we
calculate~\cite{Julienne94}
\begin{equation}
        P_{pe}(J') = |S_{pe}(\varepsilon>\Delta,\l'',B=0,\Delta,I=0)|^2.
        \label{Pep}
\end{equation}
Here $\varepsilon$ is taken above the threshold energy $\Delta$
where the $e$ channel becomes open and $S_{pe}$ is defined.

Our numerical calculations show, as expected, that $P_{pe}$ is
independent of $\varepsilon$ over a wide range, typically of
$\varepsilon/k_B$ from 0.3 mK to 300 mK at low $J'$ and 3 mK to 300 mK
at high $J'$, and also independent of $\Delta$ in our small range of
detuning.  The Landau-Zener interpretation of $P_{pe}$ in
Eq.~(\ref{LZ}) leads us to expect that $P_{pe}$ will vary with $J'$.
This variation should be stronger for the outer $^1\Pi_g$-$^3\Sigma_g$
crossing than for the inner $^1\Sigma_u$-$^3\Pi_u$ crossing.  For the
latter crossing, our calculations do give $P_{pe}$ values which vary
slowly with $J'$.  We calculate $P_{pe}(J'=1)$ to be 0.024, 0.44,
0.31, 0.34 and 0.64 for Mg, Ca, Sr, Ba, and Yb respectively.  These
probabilities are all large (order unity) except for the case of Mg.
This qualitative conclusion is likely to be robust, even though our
model calculations are only quite approximate.

\begin{figure}[htb]
%\vspace{7cm}
\noindent\centerline{
\psfig{width=85mm,figure=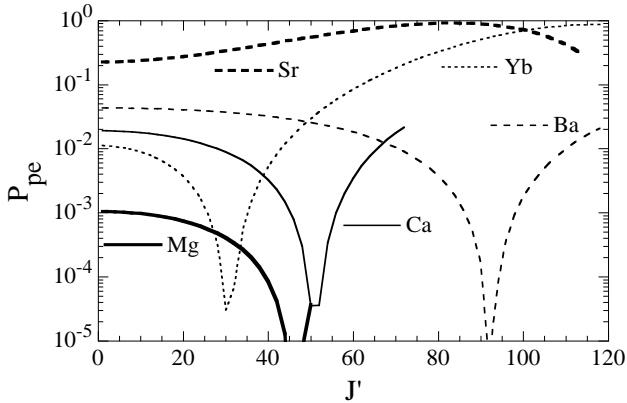}
} \caption[f4]{Calculated probabilities $P_{pe}(J')$ versus $J'$
for the ${^1}\Pi_g$-$^3\Sigma_g$ SC crossing.
\label{PepFig}}
\end{figure}

In contrast to our results for the $^1\Sigma_u$-$^3\Pi_u$ crossing,
Fig.~\ref{PepFig} shows that the calculated $P_{pe}(J')$ values
for the outer $^1\Pi_g$-$^3\Sigma_g$ crossing indeed depend much more
strongly on $J'$.  A test of the Landau-Zener formula for Mg shows
that the result of Eq.~(\ref{LZ}) is indistinguishable from the
calculated line on the figure.  The qualitative feature of a dip in
$P_{pe}$ as it goes near zero for some $J'$ is associated with the
phase factor in the LZ formula, Eq.~(\ref{LZ2}).  Since the specific
$J'$-range where this dip occurs is sensitive to the potentials
used~\cite{Dulieu94}, our model calculations can only be a qualitative
guide even for Mg.  The relative values for the other species are
also only qualitative guides, since other curve crossings are also
involved.  In any case, Sr is likely to have a large (order unity),
perhaps the largest, $P_{pe}$ for the ${^1}\Pi_g$ SC process.

We can use $P_{pg}$ and $P_{pe}$ from the close coupling calculations
to divide out the inner zone probability so as to define a numerical
excitation-transfer function from the long-range
region~\cite{Julienne94}
\begin{equation}
      J_{eg}(\varepsilon,\l'',B,\Delta,I)=\frac{P_{pg}(\varepsilon,\l'',B,
      \Delta,I)}{P_{pe}(J')} ,
      \label{JegDef}
\end{equation}
where $J'=\l''+B$.  $J_{eg}$ may be interpreted as the probability of
reaching the short-range region due to optical excitation at
long-range and propagation to short-range, including return after
multiple vibrations across the short-range well.  This interpretation
follows from the fact that one gets the total trap loss probability
$P_{pg}(\varepsilon,\l'',B,\Delta,I)$ by multiplying
$J_{eg}(\varepsilon,\l'',B,\Delta,I)$ by the probability
$P_{pe}(J')$ in Eq.~(\ref{Pep}) of a trap loss event in a single
complete cycle across the well.

Using Eq.~(\ref{JegDef}), we can define an excitation-transfer rate
coefficient $\kappa(\Delta,\varepsilon)$
\begin{equation}
        \kappa(\Delta,\varepsilon)=\frac{k_B T_\varepsilon}
        {hQ_{T_\varepsilon}}
        \sum_{l''_{even},B}(2l''+1)J_{ge}(\varepsilon,\l'',B,\Delta,I) .
        \label{kappa2}
\end{equation}
This rate coefficient $\kappa(\Delta,\varepsilon)$ is related to the
ordinary rate coefficient $K(\Delta,\varepsilon)$ in
Eq.~(\ref{sumrate}) through a mean inner zone probability $\langle
P_{pe}(\varepsilon) \rangle$, which we can {\it define} by the relation
\begin{equation}
        K(\Delta,\varepsilon)=\langle P_{pe}(\varepsilon)\rangle
        \kappa(\Delta,\varepsilon) .
        \label{kappa1}
\end{equation}
Clearly, we can also define a thermal average $\kappa(\Delta,T)$
analogous to that in Eq.~(\ref{KT}), and define a thermal
average $\langle P_{pe}(T)\rangle = K(\Delta,T)/\kappa(\Delta,T)$.

The usefulness of the factorization in Eq.~(\ref{factor1}) is that it
allows us to define an excitation-transfer rate coefficient $\kappa$
from which the inner zone SC probability has been removed (however,
see the discussion in Section \ref{ResDecay} below about how a large
$P_{pe}(J')$ may affect the width of resonance features).  We can
predict much more confidently the properties of the long-range
excitation and vibration than we can the short-range SC probabilities.
Thus, once we have a better knowledge of these short-range
probabilities, either through measurements or through better
theoretical knowledge of potential curves and couplings, we can
multiply our $\kappa$ coefficients by $\langle P_{pe}(T)\rangle$ to
get the SC rate coefficients.

\subsection{Limiting cases of the excitation-transfer probability}
\label{LimitingCases}

The attractive molecular potentials support molecular vibrational
levels with vibrational quantum numbers $v$, as described in Section
\ref{LongRange}.  We can find simple analytic expressions for the
excitation-transfer function $J_{eg}=J_{ee} P_{eg}$ factored according
to Eq.~(\ref{factor2}) for two limiting cases: (1) strongly
overlapping resonances where the probability is large for spontaneous
decay during a single vibrational cycle, i.e., the level width
is larger than the level spacing, and (2) non-overlapping, or
isolated, resonances, where many vibrations occur during a vibrational
decay lifetime, i.e., the level width is much smaller than the
level spacing.  For Group II species, ${^1}\Sigma_u$ transitions at
small detuning tend to be of the former type, but never become fully
isolated in the detuning range we consider.  On the other hand,
${^1}\Pi_g$ transitions tend to be of the latter type unless the
detuning is very small or the SC probability is very large.

\subsubsection{Small detuning and fast spontaneous decay }

The quantum mechanical theory of the first limiting case for trap loss
for small detuning and fast radiative decay has been worked out in
detail in Refs.~\cite{Julienne94,Boesten93,Suominen94,Suominen98}, where:
\begin{equation}
        J_{eg}(\varepsilon,\l'',B,\Delta,I) = J_{ee}(\varepsilon,\l'',B,\Delta)
        P_{eg}(\varepsilon,\l'',B,\Delta,I) .
        \label{factor3}
\end{equation}
The factor
\begin{eqnarray}
        P_{eg}(\varepsilon,\l'',B,\Delta,I) &=& 1 - e^{-2 \pi \Lambda} ,
        \nonumber \\
        \Lambda &=& |V_{eg}(R_C)|^2 /(\hbar v_C D_C) ,
	       \label{factor4}
\end{eqnarray}
where $v_C$ and $D_C$ are the speed and slope difference at
the Condon point, represents the Landau-Zener probability of
excitation from the ground state to the excited state in a one-way
passage through the Condon point at $R_C$.  In this limit, radiative
decay is faster than the vibrational period ($\Gamma_v \gg h \nu_v$),
there are no multiple vibrations, and the excited state transfer
factor
\begin{eqnarray}
        J_{ee}(\varepsilon,\l'',B,\Delta) &=&e^{-a_{out}} \ll 1 ,\nonumber \\
        a_{out}&=&\lambdabar\int_{u_C}^{u_{p}}{du\frac{\Gamma(u)}{\hbar v(u)}}
\end{eqnarray}
represents the probability of survival along the classical trajectory
from the Condon point of excitation to the point $R_p$ of inner zone
curve crossing; $v(u)= \left\{2[\varepsilon -V_e(u)-\Delta]
/\mu\right\}^{1/2}$ is the local classical speed.

Note that the Landau-Zener expression for $P_{eg}$ in
Eq.~(\ref{factor4}) does not have the proper Wigner law threshold
behavior, since $J_{eg}$ should be proportional to $k$ at low
collision energy.  However, our numerical $J_{eg}$ will have the
proper Wigner law form.  Note also that the quantum mechanical
calculations in Ref.~\cite{Julienne94,Suominen98,Boesten93,Suominen94}
support this semiclassical picture of localized excitation at the
Condon point, not the delocalized excitation picture of the
Gallagher-Pritchard (GP) model~\cite{GP}, which for small detuning
predicts a dominant contribution to trap loss from off-resonant
excitation at distances much less than $R_C$.  The GP model also does
not satisfy the Wigner law at low $T$.  We defer detailed comparisons
with semiclassical theories to a future publication.

\subsubsection{Non-overlapping resonances}\label{LargeDet}

The second limiting case is that of non-overlapping vibrational
resonances, that is, the spacing $h\nu_v$ [see Eq.~(\ref{LB2})] between
vibrational levels $v$ is much larger than their total width
$\Gamma_v$.  This is typical of large detuning.  Then $|S_{pg}|^2$ is
given by an isolated Breit-Wigner resonance scattering formula for
photoassociation lines~\cite{Napolitano94,Bohn99}:
\begin{equation}
        P_{pg} = {\Gamma_{vp} \Gamma_{vg} \over \left[ \varepsilon -
        (E_v + s_v) \right]^2 + \left( {\Gamma_v / 2}
        \right)^2 } .
        \label{BWres}
\end{equation}
Here $E_v=\Delta-\varepsilon_v$ is the detuning-dependent position of
the vibrational level in the molecule-field picture relative to the
ground state separated atom energy (when $\Delta=\varepsilon_v$, then
$E_v=0$ and the vibrational level is in exact resonance with colliding
atoms with zero kinetic energy), $s_v$ is a level shift due to the
laser-induced coupling, and the total width $\Gamma_v = \Gamma_{vp} +
\Gamma_{vg} +\Gamma_{v,rad}$ is the sum of the decay widths into the
probe ($\Gamma_{vp}$) and ground state ($\Gamma_{vg}$) channels and
the radiative decay rate ($\Gamma_{v,rad}$).  In the weak decay limit
($\Gamma_v \ll h\nu_v$), we can write the Fermi golden rule decay
widths as~\cite{Julienne84,Julienne86}
\begin{equation}
       \Gamma_{vi}=2 \pi
       |\langle v|V_{vi}|\varepsilon,\l\rangle|^2 = \hbar \nu_v P_{vi},
       \label{mqdt}
\end{equation}
where $i=$ $g$ or $p$, $\l$ is the partial wave for channel $i$, and
$P_{vi}$ represents the probability of decay during a {\it single
cycle} of vibration from level $v$ to channel $i$.  For the SC
process, $P_{vp}$ is a very weak function of energy as long as the
detuning is not too large, and we can take $P_{vp}=P_{pe}$, where
$P_{pe}$ is the energy-independent SC probability discussed in
Sec.~\ref{Factorization}.  In the weak-field limit, $\Gamma_{vg}$ is
very small in relation to $\Gamma_{v,rad}$, and we can ignore it (that
is, there is no power broadening).

Using Eq.~(\ref{mqdt}) in Eq.~(\ref{BWres}), we get the factorization in
Eq.~(\ref{factor3}) with the resonant-enhanced transfer function
\begin{equation}
       J_{ee} = {(\hbar \nu_v)^2 \over \left[ \varepsilon -
        (E_v + s_v) \right]^2 + \left( {\Gamma_v / 2}
        \right)^2 } .
        \label{Jee}
\end{equation}
If we use the {\it reflection approximation} for the Franck-Condon
factor in $\Gamma_{vg}$~\cite{Julienne96,Boisseau00,Bohn99}, then
\begin{equation}
        P_{eg} = 4\pi^2 |V_{eg}(R_C)|^2 \frac{1}{D_C}
        |\phi_g(\varepsilon,\l'',R_C)|^2 .
        \label{RefApprox}
\end{equation}
Equation~(\ref{RefApprox}) may be used throughout the whole cold
collision domain (mK to nK).  It satisfies the the correct Wigner
threshold law behavior at low energies because of the $|\phi_g|^2$
factor.  In the $s$-wave limit for low temperature, we may take the
asymptotic form of the ground state wavefunction and obtain:
\begin{equation}
        P_{eg} = 16\pi^2 \frac{|V_{eg}(R_C)|^2}{h v_{\infty} D_C}
        \sin^2 k(R_C-A_0) .
        \label{sRefApprox}
\end{equation}
This looks just like the Landau-Zener result in Eq.~(\ref{LZ}), except
that the asymptotic speed $v_{\infty}$ appears in the denominator instead
of $v_C$~\cite{Julienne96}, and the correct quantum phase appears in the
sine factor instead of a semiclassical phase.

\subsubsection{Contributions to the linewidths }\label{ResDecay}

The expression, Eq.~(\ref{factor3}), for $J_{eg}$ in the limit of
small detuning and fast decay does not depend in any way on the
short-range SC probability.  However, in the expression for $J_{eg}$
in the isolated resonance limit, the width $\Gamma_{v}$ in the
$J_{ee}$ factor, Eq.~(\ref{Jee}), does depend on $P_{pe}$ through the
contribution of $\Gamma_{vp}=\hbar \nu_v P_{pe}$.  As long as
$\Gamma_{vp}$ is small compared to $\Gamma_{v,rad}$, the total width
$\Gamma_v$ is determined primarily by $\Gamma_{v,rad}$, and the shape
of trap loss spectral lines will still be nearly independent of
$P_{pe}$.  However, if $\Gamma_{vp}$ makes a significant contribution
to the total width, the long-range excitation-transfer function
$J_{eg}$ will show additional broadening dependent on the magnitude of
$P_{pe}$.

The total radiative decay width $\Gamma_{v,rad}$ can be calculated
from the long-range form of the decay rates in
Eqs.~(\ref{mollinewidth}), using the excellent semiclassical
approximation~\cite{Tell84}, $\langle v|\Gamma(u)|v\rangle = \nu_v
\oint_{v} (\Gamma(u)/v(u)) \lambdabar du$, where the semiclassical
integral is over a complete vibrational cycle.  When $R_C <
\lambdabar$, we can use the lead term in the expansion of $\Gamma(u)$
in $u$ in Eqs.~(\ref{mollinewidth}), so that
\begin{eqnarray}
     \Gamma_{v,rad}({^1}\Sigma_u) &=& 2{\Gamma_{\mrm at}} = {\mrm constant} ,\\
     \Gamma_{v,rad}({^1}\Pi_g) &=& {\Gamma_{\mrm at}} \frac{\pi}{20 a_3}
      u_C^2 = 0.701
      \Gamma_C({^1}\Pi_g) ,
      \label{RadWidths}
\end{eqnarray}
where $a_3$ is defined after Eq.~(\ref{LB1}) and
${\Gamma{_C}}({^1}\Pi_g)$ is evaluated at the outer turning point of
the vibration, which is almost the same as the Condon point.

For the detuning range we consider, the radiative width of
${^1}\Sigma_u$ levels, $2\Gamma_{\mrm at}$, is much larger than
$\Gamma_{vp}$, which can be calculated from Eq.~(\ref{mqdt}) using the
probabilities listed in Sec.~\ref{Factorization}.  Thus, $\Gamma_v
\approx \Gamma_{v,rad}$ so that the {\it shape} of ${^1}\Sigma_u$
features (that is, their spacings and widths) should be
well-determined in our calculations.

\begin{figure}[htb]
%\vspace{7cm}
\noindent\centerline{
\psfig{width=85mm,figure=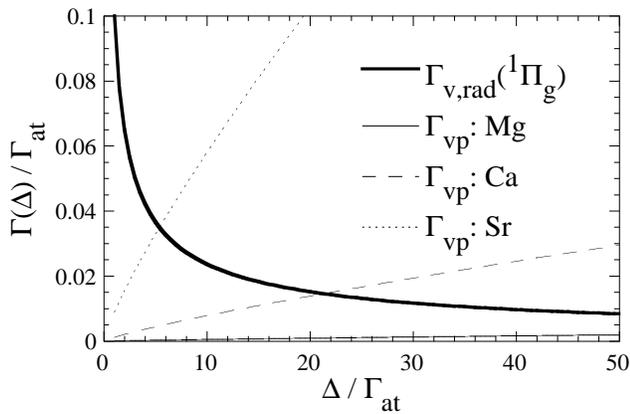}
} \caption[f5]{Radiative width $\Gamma_{v,rad}$ for Mg and widths
$\Gamma_{vp}$ for Mg, Ca and Sr versus $\Delta$ for the
${^1}\Pi_g$-$^3\Sigma_g$ SC crossing.
\label{Width}}
\end{figure}

Figure~\ref{Width} shows $\Gamma_{v,rad}$ and $\Gamma_{vp}$ for
${^1}\Pi_g$ features for Mg, Ca, and Sr.  In our detuning range,
$\Gamma_{v,rad} \gg \Gamma_{vp}$ for Mg.  Thus, the shape of Mg
${^1}\Pi_g$ features should also be well-determined in our calculations.
On the other hand, for Ca and Sr, the $\Gamma_{vp}$ is larger due to the
larger $P_{pe}$. $\Gamma_{vp}$ increases as $\Delta^{5/6}$ due to the
$\nu_v$ factor in Eq.~(\ref{mqdt}), and becomes larger than
$\Gamma_{v,rad}$ near $\Delta=5\Gamma_{\mrm at}$ in our model for Sr and
near 20$\Gamma_{\mrm at}$ for Ca.  Thus, we can expect predissociation
broadening of ${^1}\Pi_g$ features to become observable for Ca or Sr at
relatively small detunings. Measurements of such widths could lead to
experimental information about $P_{pe}$ for the ${^1}\Pi_g$ state.  On
the other hand, our calculated model line shapes should only be viewed as
a qualitative guide in a region where $\Gamma_{vp} \gg \Gamma_{v,rad}$.

\subsection{Radiative escape calculations}\label{REcalculations}

The calculation of the RE trap loss rate coefficient follows the
factorization procedure in Eq.~(\ref{factor1}) as for the SC process.
The RE loss is not due to a single curve crossing, but rather to
excited state emission from the distance range $R<R_p=\lambdabar u_p$,
where $u_p$ is the point for which a kinetic-energy increase of 1 K
for the atom pair has been gained after excitation (the 1 K is
arbitrary--we only choose it to represent a ``standard'' loss energy).
Clearly, RE can only be significant for the $^1\Sigma_u$ state,
because of the negligible short-range emission from the $^1\Pi_g$
state.  We calculate the total probability of radiative escape,
$P_{pe}=P_{\mrm decay}$ during a complete cycle of vibration across
the region $u<u_p$ by integrating along the classical trajectory:
\begin{eqnarray}
P_{\mrm
decay}(\varepsilon,J',\Delta)&=&1-\exp(-a),\nonumber \\
a&=&2\lambdabar  \int_{u_p}^{u_{in}}{du\frac{\Gamma(u)}{\hbar v(u)}}.
\end{eqnarray}
\noindent $P_{\mrm decay}$ depends only weakly on $\Delta$, $J'$.
Variations with $\varepsilon$ at the highest collision energies also play
a role when calculating the thermally averaged rate.  The main
contribution to $P_{\mrm decay}$ comes from the long-range region where
the potential is determined by its analytic long-range form.

The $P_{\mrm decay}$ probability is insensitive to collision energy
and detuning.  In the detuning range $\Delta/\Gamma_{\mrm at}$ from 1
to 50 and for a collision energy of $k_B T_D$, we find that $P_{\mrm
decay}$ ranges from 0.157 to 0.144 for Mg, 0.103 to 0.100 for Ca,
0.147 to 0.142 for Sr, 0.113 to 0.110 for Ba, and 0.149 to 0.145 for
Yb.  These hardly change at all at a collision energy of $k_B
T_D/1000$, for example, changing to 0.105 to 0.100 for Ca.

We have also used a calculation with an artificial probe state
crossing the excited state potential at $R_p$ ($R_p \approx 150$ a$_0$ for
our Mg model for $J'=1$), as described in Sec.~\ref{MgModel}, to
calculate the excitation-transfer function $J_{eg}$ appropriate to the
RE process.  We find, as expected, that the numerical $J_{eg}$
function calculated this way is very nearly the same as the one
calculated using the SC $R_p$ at much shorter range.  For our detuning
range the radiative contribution to the total width $\Gamma_v$ of
$^1\Sigma_u$ levels is much larger than contribution due to
predissociation to the SC channel.

We expect that our radiative escape trap loss calculations
are reliable in magnitude, since only long-range properties are
relevant in determining both $J_{eg}$ and $P_{\mrm decay}$.
Therefore, in the next Section we can confidently give absolute
magnitudes for the RE contribution to the total trap loss rate
coefficient $K(\Delta,T)$ for all alkaline earth atoms we study here.

\section{Results}\label{resultsMg}

\subsection{Trap loss for Mg at $T=T_D$}\label{MgT_D}

\begin{figure}[htb]
%\vspace{7cm}
\noindent\centerline{
\psfig{width=85mm,figure=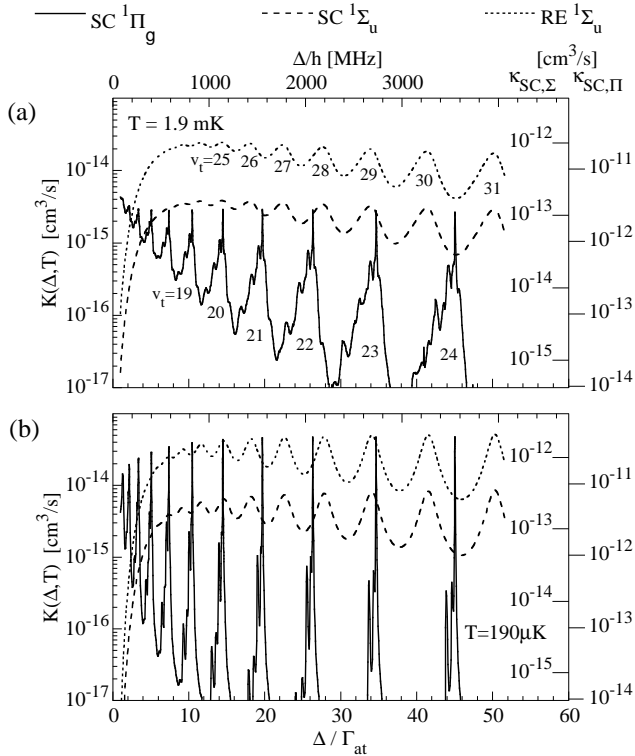}
} \caption[f6]{Contributions from the ${^1}\Sigma_u$ RE and
${^1}\Sigma_u$ and ${^1}\Pi_g$ SC processes to the thermally averaged
loss rate coefficient $K(\Delta,T)$ at (a) 1.9 mK and (b) 190 $\mu$K
as a function of laser detuning $\Delta$ for Mg at a standard laser
intensity $I=1$ mW/cm$^2$.  The scales for the excitation-transfer
coefficients, $\kappa(\Delta,T)$, for the SC processes are indicated
by the vertical axes to the right.  The vibrational quantum numbers from
the top of the potential, $v_t$, are indicated for ${^1}\Sigma_u$ and
${^1}\Pi_g$ features.
\label{MgEA}}
\end{figure}

\begin{figure}[htb]
%\vspace{7cm}
\noindent\centerline{ \psfig{width=85mm,figure=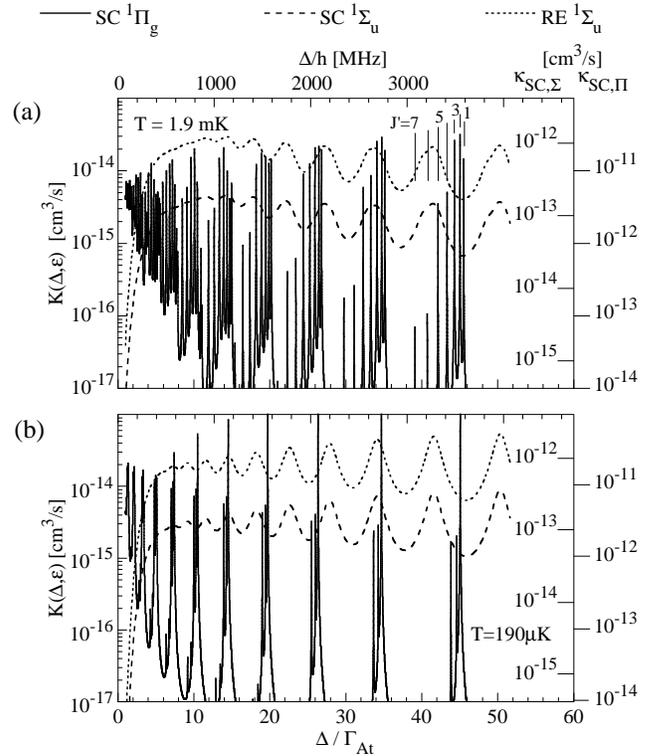} }
\caption[f7]{Contributions from the ${^1}\Sigma_u$ RE and
${^1}\Sigma_u$ and ${^1}\Pi_g$ SC processes to the loss rate
coefficients $K(\Delta,\varepsilon)$ at a fixed collision energy (a)
$\varepsilon = k_B$(1.9 mK) (b) $\varepsilon = k_B$(190 $\mu$K) as a
function of laser detuning $\Delta$ for Mg at laser intensity $I=1$
mW/cm$^2$.  $K(\Delta,\varepsilon)$ is a sum over partial waves and
branches for $\varepsilon=k_BT$.  The corresponding excitation-
transfer coefficients, $\kappa(\Delta,\varepsilon)$, are indicated by
the vertical axis to the right.  Excited state rotational quantum 
numbers are indicated for the $v_t=24$ ${^1}\Pi_g$ feature in (a).
\label{MgSum}}
\end{figure}

\begin{figure}[htb]
%\vspace{7cm}
\noindent\centerline{
\psfig{width=70mm,figure=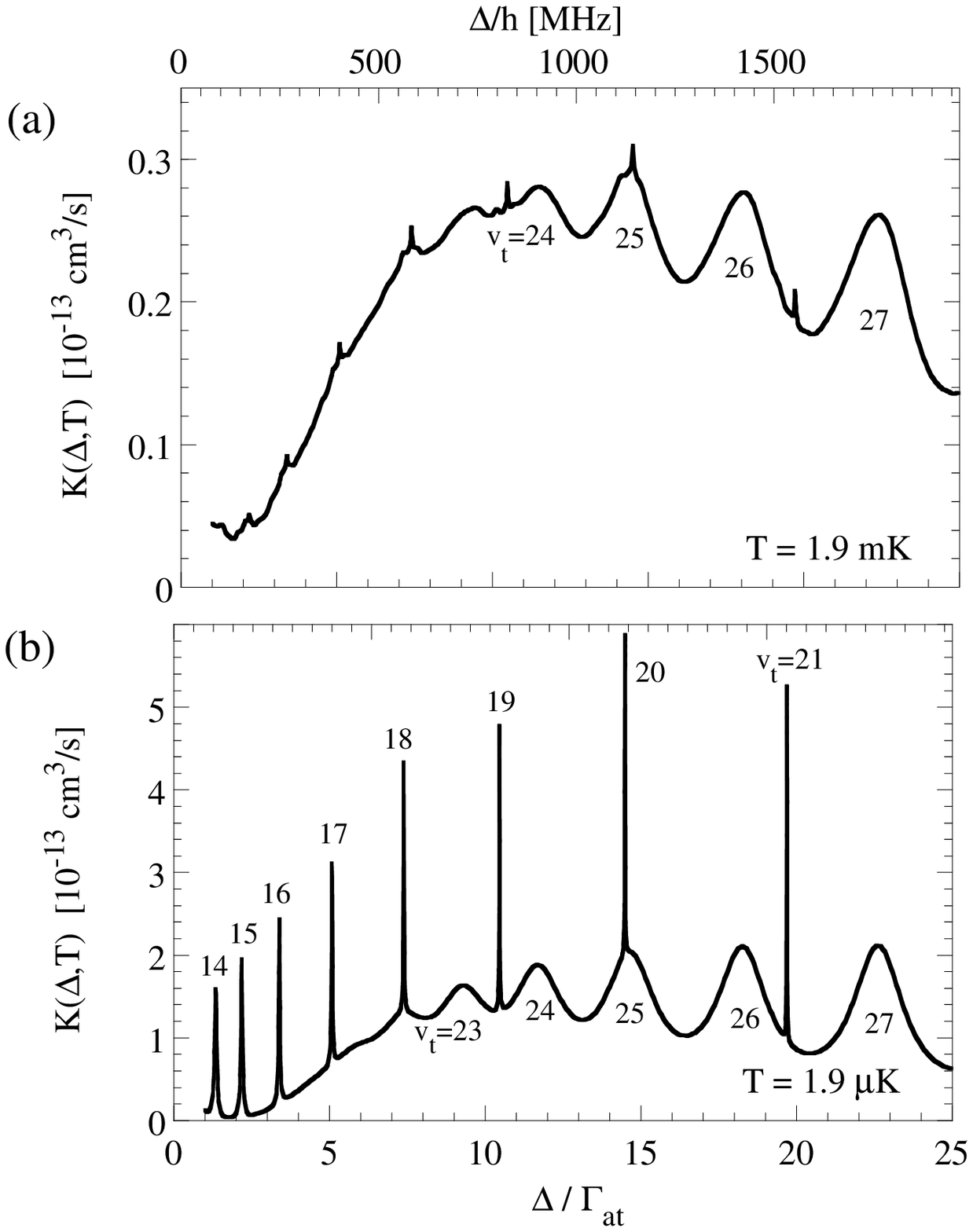}
} \caption[f8]{Total thermally averaged Mg spectrum, $K(\Delta,T)$
summed over all RE and SC contributions, on a linear scale.  (a) At
$T_D=$1.9 mK, (b) In the $s$-wave limit at $T= 1.9\ \mu$K.  The 
vibrational quantum numbers $v_t$ are indicated for the ${^1}\Sigma_u$ 
and ${^1}\Pi_g$ features.  Only excited $J'=1$ levels contribute 
R-branch transition from $s$-waves in panel (b).
\label{LinearKTot}}
\end{figure}

Our calculated results for $T_D = 1.9$ mK for Mg collisions are shown
in Figs.~\ref{MgEA}(a), \ref{MgSum}(a), and \ref{LinearKTot}(a).
These results are different from the results presented in
Ref.~\cite{Machholm99}, since we have corrected some errors
we made in the sum over partial waves in that
reference~\cite{Corrections}.  Figure ~\ref{MgEA}(a) shows on a
logarithmic scale the separate contributions of each SC or RE process
to the thermally averaged rate constant $K(\Delta,T)$ from
Eq.~(\ref{thermalrate}), whereas Fig.~\ref{MgSum}(a) shows the
corresponding results for $K(\Delta,\varepsilon)$ at a single
collision energy $\varepsilon$.  Figure \ref{LinearKTot}(a) shows on a
linear scale the sum of contributions from all loss processes, and
shows what one might expect to see in a laboratory spectrum.

The dominant loss process for Mg at 1.9 mK is due to RE from the
${^1}\Sigma_u$ state.  The spectra for ${^1}\Sigma_u$ RE and SC
processes have the same shape, since they have the same
excitation-transfer function $\kappa$.  The RE and SC processes differ
only by a multiplicative factor that is nearly independent of
$\Delta$, due to the different $P_{pe}$ factors for RE and SC. The
${^1}\Sigma_u$ RE probability only varies by 0.157 to 0.144 from
detunings of 1 to 50 $\Gamma_{\mrm at}$, whereas the ${^1}\Sigma_u$ SC
probability is constant over this range.  The ${^1}\Sigma_u$ spectra
in Figs.~\ref{MgEA}(a) and \ref{MgSum}(a) are nearly the same, since
the broad features do not change much upon thermal averaging.  The
rate coefficient becomes very small as detuning decreases below 2 or 3
$\Gamma_{\mrm at}$.  This is because $J_{ee} \ll 1$ for very small
detuning due to spontaneous emission during the long-range approach of
the two atoms.  Spontaneous emission losses become small for detunings
larger than around 10 $\Gamma_{\mrm at}$, and vibrationally resolved,
but rotationally unresolved, photoassociation structure begins to
develop as detuning increases.  This occurs as the spacing between
adjacent ${^1}\Sigma_u$ vibrational levels from Eq.~(\ref{LB2})
becomes larger than the radiative decay width.  Several rotational
features with different $J'$ may contribute to each of the broad
photoassociation resonances, with the range of $J'$ depending on
detuning.  Each individual ${^1}\Sigma_u$ rotational line has a width
on the order of $(2\Gamma_{\mrm at}+k_B T)/h \approx$ 200 MHz.  There
is negligible predissociation broadening due to SC processes in this
region of the spectrum.

Using Eq.~(\ref{LB1}) in Section \ref{LongRange}, the vibrational
quantum number $v_t$, as counted down from the top of the potential at
the dissociation limit, can be given for the resolved, or partially
resolved, features in the trap loss spectrum.  We define $v_t$ to be
$v_D - v$ rounded up to the next integer.  Each integer $v_t$ value
defines an energy range which contains only one vibrational level for
a given $J$.  The $v_t$ quantum numbers for ${^1}\Sigma_u$ and
${^1}\Pi_g$ features are indicated on Fig.~\ref{MgSum}.  Note that
there are many levels (not calculated) within the range
$\Delta/\Gamma_{\mrm at} < 1$, a range where Eq.~(\ref{LB1}) is not
meaningful due to retardation effects on the potential.  The broad
${^1}\Sigma_u$ features provide an example of overlapping resonances,
analogous to those treated by Bell and Seaton~\cite{Bell85} for the
case of dielectronic recombination where the spacing between
collisional resonance levels becomes less than their radiative decay
width.

The contribution to $K(\Delta,T)$ from the ${^1}\Pi_g$ SC process
shows much sharper vibrational structure than the corresponding
${^1}\Sigma_u$ spectrum.  This is because of the small radiative
widths of the ${^1}\Pi_g$ levels, which become even smaller as
$\Delta$ increases.  The individual contribution from a number of
narrow rotational levels is evident in Fig.~\ref{MgSum}(a).  Figure
\ref{MgEA}(a) shows that this ${^1}\Pi_g$ structure even survives
thermal averaging.  Figure \ref{LinearKTot}(a) shows that sharp
${^1}\Pi_g$ features can even survive thermal averaging at 1.9 mK,
although such features are quite weak for Mg and would be hard to see
(However, see below for Ca and Ba, where such features might be
observable).  We find that there are sharp $J'=1$ ${^1}\Pi_g$ features
due to $s$-wave collisions that can be much narrower than $k_B T$
(which is about 40 MHz at 1.9 mK), whereas features due to $\l''>0$
collisions have widths on the order of $k_B T$.  This $s$-wave
behavior is evident in our numerical calculations, but can be easily
explained in terms of the analytic behavior of the isolated line
shapes using Eqs.~(\ref{BWres}), (\ref{mqdt}), and (\ref{RefApprox}).
We will discuss this $s$-wave resonance narrowing feature in more
detail elsewhere.

Figures \ref{MgEA} and \ref{LinearKTot} both show that at very small
detuning, on the order of 1 or 2 $\Gamma_{\mrm at}$, the trap loss is
dominated by SC due to the ${^1}\Pi_g$ state.  The increasing
radiative transition probability as detuning decreases, and the near
absence of spontaneous emission losses for the weakly emitting state,
ensures that the ${^1}\Pi_g$ contribution to trap loss must be
dominant at very small $\Delta$.  We will show in the next section
that this is even more important for the heavier species.  Our
conclusion concerning the role of the ${^1}\Pi_g$ state at small
$\Delta$ agrees with the findings of
Refs.~\cite{Gallagher99,Machholm99}.

Figures \ref{MgEA}(b) and \ref{MgSum}(b) show the contributions to SC
and RE processes for Mg at 190 $\mu$K. The broad ${^1}\Sigma_u$
features are not very sensitive to changing the temperature.  They
narrow slightly at the lower temperature.  However, the ${^1}\Pi_g$
features simplify and clearly have contributions from fewer partial
waves.  The effect of thermal averaging on ${^1}\Pi_g$ features is to
cause some broadening, with consequent decrease in peak height.

\begin{figure}[htb]
%\vspace{7cm}
\noindent\centerline{
\psfig{width=70mm,figure=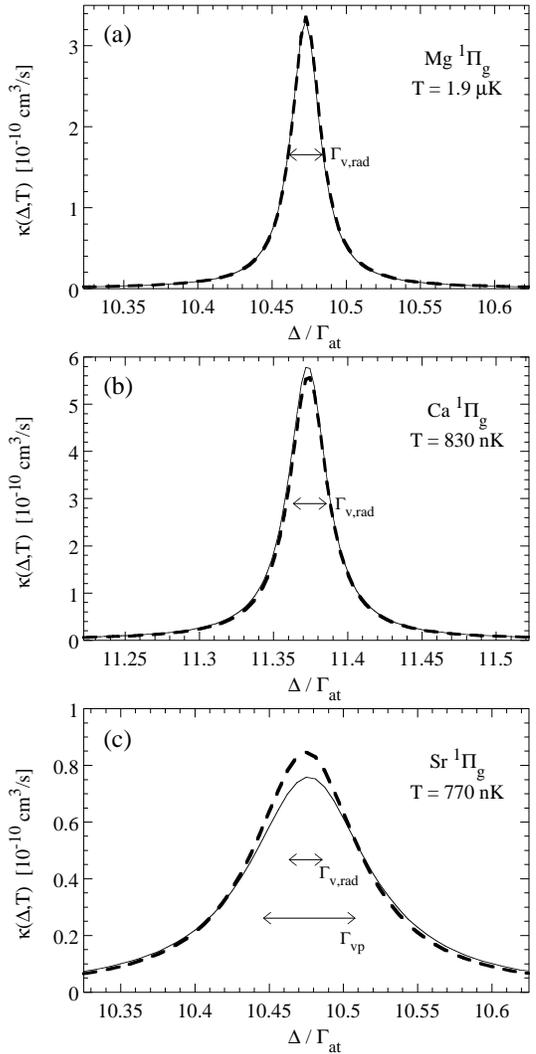}
} \caption[f9]{Single ${^1}\Pi_g$ vibrational feature in the vicinity
of $\Delta \approx 10 \Gamma_{\mrm at}$ for (a) Mg, (b) Ca, and (c)
Sr.  The figure shows the quality of the isolated resonance
approximation for the excitation-transfer line shape
$\kappa(\Delta,T)$.  The solid line is the complex potential numerical
calculation, and the dashed line is the analytic line shape based on
Eqs.~(\ref{factor3}), (\ref{Jee}), and (\ref{sRefApprox}).}
\label{SingleRes}
\end{figure}

\subsection{Trap loss for Mg near 1 $\mu$K}\label{lowTMg}

Figure \ref{LinearKTot}(b) shows $K(\Delta,T)$ summed over all
components at the extremely cold temperature of 1.9 $\mu$K. This is
deeply in the Wigner law domain, where only $s$-wave collisions
contribute to the spectrum, and the rate constant $K(\Delta,T)$
becomes independent of $T$~\cite{Weiner99}.  The broad ${^1}\Sigma_u$
features are similar to the ones at higher temperature, but are due
only to absorption by a single R branch line from $\l''=0$ to a $J'=1$
${^1}\Sigma_u$ level.  The only significant broadening is due to
radiative decay.  On the other hand, the ${^1}\Pi_g$ features, also
due to a single R branch line from $\l''=0$ to a $J'=1$ ${^1}\Pi_g$
level, become prominent sharp features in the spectrum, having widths
on the order of a few MHz due to radiative decay.  Even the level near
1 $\Gamma_{\mrm at}$ detuning is quite sharp and isolated.  Section
\ref{ShortRange} discusses why we expect to get the vibrational spacings
right, although we do not expect to predict correctly the actual position
of levels, which depend on an unknown phase due to the short-range
${^1}\Pi_g$ potential.

Figure \ref{SingleRes}(a) shows the excellent quality of the isolated
resonance approximation for a Mg ${^1}\Pi_g$ $s$-wave absorption
feature due to a single vibrational level.  This approximation should
be good in this case, since the mean vibrational spacing near this
level is 280 MHz, which is much larger than the width.  The Figure
compares the numerical line shape with that calculated using the
isolated resonance formulas discussed in Section \ref{LimitingCases}.
The analytic formula calculates the factors in Eqs.~(\ref{factor3}),
which are used in Eq.~(\ref{kappa2}), by making the isolated resonance
approximation, Eq.~(\ref{Jee}), and the reflection approximation,
Eq.~(\ref{sRefApprox}).  The linewidth in the denominator of
Eq.~(\ref{Jee}), calculated to be 1.6 MHz from Eq.~(\ref{RadWidths}),
is almost entirely due to weak spontaneous decay of this ${^1}\Pi_g$
level, as discussed Section \ref{LimitingCases} in relation to
Fig.~\ref{Width}.  Any broadening due to thermal averaging is
negligible, since $k_B T/h = $ 0.04 MHz.

We have verified that our thermal spectrum at relatively high
temperature, 1.9 mK, is to a good approximation independent of the
choice of ground state potential, as discussed in Section
\ref{GroundState}.  This is because of the need to sum over several
partial waves, for which the phase of the ground state wavefunction
varies by more than $\pi$.  In addition, the need to average over a
range of collision energies also contributes a range of phase
variation to the ground state wavefunction.

\begin{figure}[htb]
%\vspace{7cm}
\noindent\centerline{
\psfig{width=85mm,figure=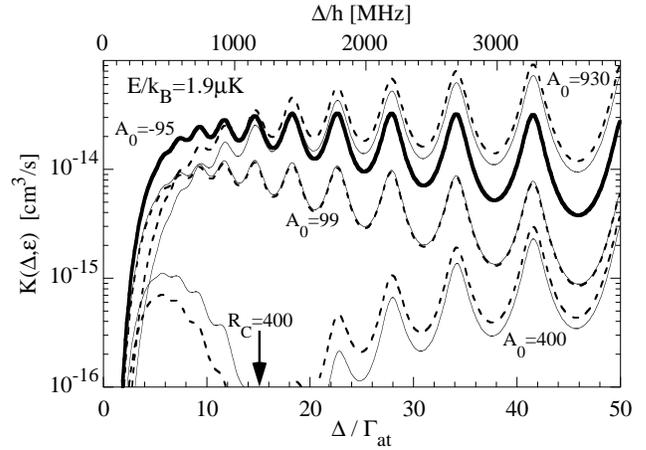}
} \caption[f10]{Scaling with different scattering lengths of
$K(\Delta,\varepsilon)$ for the ${^1}\Sigma_u$ transition in Mg.  The bold
solid line shows the numerically calculated $K(\Delta,\varepsilon)$ at
$\varepsilon = k_B (1.9$ $\mu$K) for the ``standard'' ground state
model potential with $A_0=-95$ a$_0$.  The other solid lines show the
calculated $K(\Delta,\varepsilon)$ for three other model potentials
with different scattering lengths of 99 a$_0$, 400 a$_0$, and 930
a$_0$.  The dashed line shows the scaled $K(\Delta,\varepsilon)$
calculated from the ``standard'' one using the scaling relation
discussed in the text.  The detuning for which the Condon point is 400
a$_0$ is indicated by the arrow.  The effect of the node in the ground
state wavefunction is evident for the $A_0=400$ a$_0$ case.
\label{ScaleScatLen}}
\end{figure}

The spectrum at very low temperature, on the other hand, is sensitive
to the phase of the ground state wavefunction, which is generally
unknown for Group II species and strongly dependent of the details of
the ground state potential.  This sensitivity is explained by the
reflection approximation in Eq.~(\ref{sRefApprox}), which shows
$P_{eg}$ is proportional to $\sin^2 k(R_C-A_0)$.  We have just seen
that the reflection approximation is excellent for isolated resonance
line shapes.  Therefore, if we know $K(\Delta,T)$ in the $s$-wave
domain for one scattering length $A_0$, and if we have a different
potential with a different scattering length $A_0'$, the $K(\Delta,T)$
for the new case can be scaled from the original one by multiplying by
the ratio $\sin^2 k(R_C-A_0')/\sin^2 k(R_C-A_0)$.  Figure
\ref{ScaleScatLen} compares this scaling (dashed lines) to numerical
calculations (solid lines) for several different model ground state
potentials with different $A_0'$.  The former are scaled from our
original calculation, for which $A_0=-95$ a$_0$.  Figure
\ref{ScaleScatLen} demonstrates that this scaling is a good
approximation, even when the scattering length is unusually large and
even for overlapping ${^1}\Sigma_u$ features.  The scaling relation is
excellent at small $\Delta$ for scattering lengths having magnitudes
up to a few times $x_0$ (defined in Sec.~\ref{GroundState} and having
a value of 36 a$_0$ for Mg).  The scaling is even a reasonable
approximation for the case where $A_0'=400$ a$_0$ and the ground state
wavefunction has a node at $R_C=A_0'$ near $\Delta/ \Gamma_{\mrm at} =
15$.  The node for the $A_0'=930$ a$_0$ case occurs for
$\Delta/\Gamma_{\mrm at} < 1$ and is off scale in
Fig.~\ref{ScaleScatLen} for the $A_0'=99$ a$_0$ case.

\section{Other alkaline earth atoms}\label{resultsOther}

\begin{figure}[htb]
%\vspace{7cm}
\noindent\centerline{
\psfig{width=85mm,figure=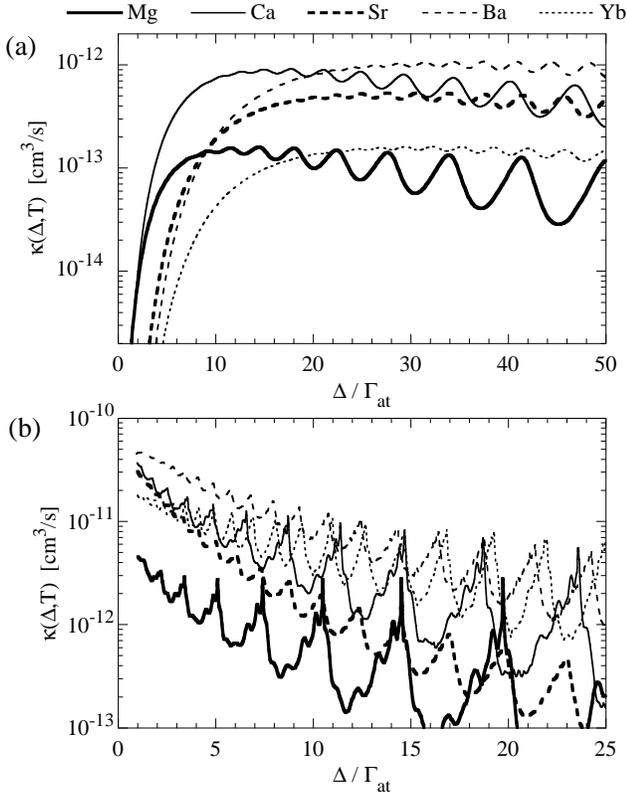}
} \caption[f11]{Excitation transfer coefficients
$\kappa(\Delta,\varepsilon)$ on a logarithmic scale for the (a)
$^1\Sigma_u$ and (b) $^1\Pi_g$ states as a function of laser detuning
$\Delta$ for Mg, Ca, Sr, Ba, and Yb at a laser intensity of $I=1$
mW/cm$^2$.
\label{KappaOthers}}
\end{figure}

\begin{figure}[htb]
%\vspace{7cm}
\noindent\centerline{
\psfig{width=85mm,figure=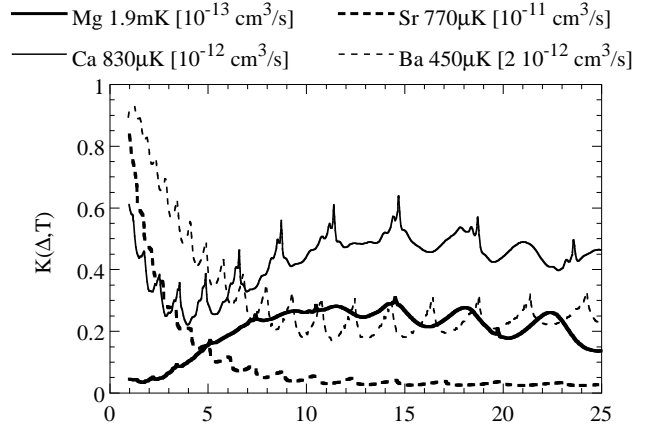}
} \caption[f12]{Spectrum $K(\Delta,T)$ summed over all RE and SC
contributions at $T_D$ for Mg, Ca, Sr, and Ba at a laser intensity of
$I=1$ mW/cm$^2$.
\label{LinearKOthersTD}}
\end{figure}

\begin{figure}[htb]
%\vspace{7cm}
\noindent\centerline{
\psfig{width=70mm,figure=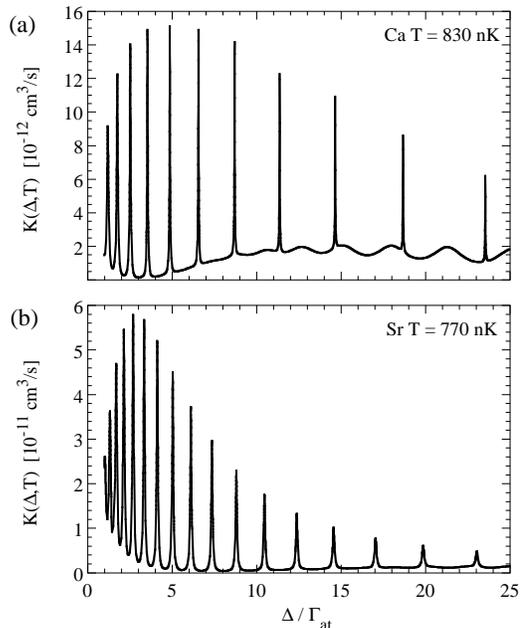}
} \caption[f13]{Spectrum $K(\Delta,T)$ summed over all RE and SC
contributions at $s$-wave domain at $T_D/1000$ for (a) Ca and (b) Sr
at a laser intensity of $I=1$ mW/cm$^2$.
\label{LinearKOthersT0}}
\end{figure}

Our calculations for the other Group II species and Yb are shown in
Figs.~\ref{SingleRes}(b), \ref{SingleRes}(c), \ref{KappaOthers},
\ref{LinearKOthersTD}, and \ref{LinearKOthersT0}.  We trust that these
model calculations, which can only provide order of magnitude
estimates for SC probabilities and predissociation contributions to
linewidths, will provide a useful qualitative guide to differences and
similarities among the various cases to guide future experiments on
these systems.  Our calculations should be fairly robust with respect
to qualitative expectations as to the different kinds of features to
expect in trap loss spectra.

Figure \ref{SingleRes}(b) shows that a very low temperature Ca
${^1}\Pi_g$ feature is very similar to the Mg one previously
discussed.  The total width is slightly larger than the radiative
width due to weak predissociation of this feature (see
Fig.~\ref{Width}).  The effect of the large predissociation width,
where $\Gamma_{vp} > \Gamma_{\mrm rad}$, is evident for the Sr feature
in Fig.~\ref{SingleRes}(c).  The isolated resonance approximation is
also beginning to fail for Sr lines because of the strong
predissociation broadening in our model with $P_{pe} = 0.22$ (see
Fig.~\ref{PepFig}).  Although our model calculation for Sr
predissociation widths should not be considered to be reliable, the
model does show that if the widths of ${^1}\Pi_g$ features like the
one in Fig.~\ref{SingleRes}(c) could be measured, the data should
allow a value to be determined for $P_{pe}$.  Since temperatures in
the nK regime have already been reported for intercombination line
cooling of Sr, it may be quite feasible to measure such widths.

Figure \ref{KappaOthers} shows the thermally averaged
excitation-transfer coefficients $\kappa(\Delta,T)$ (see
Eq.~(\ref{kappa2}) and following) for the ${^1}\Sigma_u$ and
${^1}\Pi_g$ states in these systems at $T_D$ for the ${^1}$S$\to
{^1}$P cooling transition.  In spite of the fact that the inner zone
SC probability $P_{pe}$ is divided out of the expression for
$\kappa(\Delta,T)$, there are still a number of differences among the
different species.  The differences in spacing and contrast of the
individual vibrational features that appear at larger detuning is
clearly related to the vibrational spacing, Eq.~(\ref{LB2}), which
decreases with increasing mass.  The differences in magnitude can be
qualitatively related to the scaling of the different factors that
make up $\kappa(\Delta,\varepsilon)$ in Eq.~(\ref{kappa2}).  There are
four factors that contribute to the scaling: (1) $1/Q_T \to
\mu^{-3/2}$, (2) the sum over $\l'' \to \l_{\mrm max}^2 \to \mu
d_0^{4/3}/\Delta^{2/3}$, (3) $J_{ee}({\mrm peak}) \to
(\nu_v/\Gamma_v)^2 \to \lambdabar^6 \Delta^{5/3} /(\mu d_0^{16/3})$,
and (4) $P_{eg} \to |V_{eg}(R_C)|^2/(v D_C) \to \mu^{1/2} d_0^{8/3} /
\Delta^{4/3}$.  The net scaling of the peak magnitude of
$\kappa(\Delta,T)$ thus scales approximately as $\lambdabar^6/(\mu
d_0^{4/3} \Delta^{1/3})$.  This gives scaling factors at the same
$\Delta$ of 1, 5.0, 3.4, 6.4, 1.1 for Mg, Ca, Sr, Ba, and Yb
respectively (these factors should be scaled by an additional factor
of $\lambdabar / d_0^{2/3}$ if evaluated at the same scaled detuning,
$\Delta/\Gamma_{at}$).  These scaling factors account for the relative
magnitudes of the peak $\kappa(\Delta,T)$ for the ${^1}\Sigma_u$ state
in Fig.~\ref{KappaOthers}(a) in the relatively flat region from 20 to
50 $\Delta/\Gamma_{\mrm at}$.  The scaling for the ${^1}\Pi_g$ spectra
in Fig.~\ref{KappaOthers}(b) also needs to take into account the
predissociation contribution to the width $\Gamma_v$, which was taken
to be purely radiative for the scaling of the ${^1}\Sigma_u$ spectrum
in Fig.~\ref{KappaOthers}(a).  For example, this extra predissociation
broadening lowers the peak of Sr features below those for Mg in
Fig.~\ref{KappaOthers}(b).

Figure \ref{LinearKOthersTD} shows our model thermally averaged
$K(\Delta,T)$ summed over all contributions.  With the {\it caveat}
that the relative contributions of SC processes are not likely to be
reliable in our model calculations, these model spectra show
qualitative features that one might observe in laboratory spectra.  In
particular, ${^1}\Pi_g$ SC processes make a dominant contribution to
the small detuning trap loss for $\Delta <$ a few $\Gamma_{\mrm at}$. 
This has already been discussed in
Refs.~\cite{Gallagher99,Machholm99}.  We can compare our results to
the measured $2K(\Delta,T) = 4.5(0.3)(1.1)\times 10^{-10}$ cm$^3/$s
\cite{GallagherNote} for Sr at $\Delta/\Gamma_{\mrm at}= 1.75$, $I =
60$ mW/cm$^2$, and $T \approx 4 T_D$ \cite{Gallagher99}.  Although the
effect of a strong laser field needs to be investigated for this case,
Ref.~\cite{Suominen98} suggests that near-linear scaling may apply to
small detuning trap loss even in the strong field domain (see Fig.~6
of that reference).  If we assume linear scaling with $I$, our
calculated value for $T=T_D$ at $I=1$ mW/cm$^2$ scales to a value of
$2K = 6 \times 10^{-10}$ cm$^3/$s at $I = 60$ mW/cm$^2$.  The
agreement of our very approximate model to within a factor of two with
the measured result for Sr is gratifying and lends confidence to the
usefulness of our estimates.

At the present, there are no other data on Sr or other Group II
species to which we can compare our calculations directly.  The Ca$_2$
photoassociation spectra in a 3 mK Ca MOT reported by Zinner {\it et
al.} in Ref.~\cite{Tiemann00} extend over a detuning range from about
50 to 2700 $\Gamma_{\mrm at}$, which is larger than we calculate.
They observed ${^1}\Sigma_u$ features and gave a detailed analysis of
partially resolved rotational substructure for a feature near $\Delta
= 27$ GHz $=$ 780 $\Gamma_{\mrm at}$.  The fact that the width of this
feature could be explained by a combination of natural and thermal
broadening of several rotational lines implies that predissociation
broadening makes a small contribution to the linewidth of this
feature.  If we assume that 20 MHz or less of the observed 150 MHz
feature width is due to predissociation, we would then estimate the
${^1}\Sigma_u$ SC $P_{pe} < 0.05$, which is much less than the value
0.4 estimated by our model for Ca.  Although our model should not be
extrapolated to such large detuning without careful testing, this
apparent inconsistency points out that much more detailed knowledge of
potentials and matrix elements is needed for accurate calculations.
It is an interesting fact to be explained why the apparent
predissociation rate of ${^1}\Sigma_u$ levels in Ca$_2$ is relatively
small, given the likelihood of several curve crossings with moderately
large matrix elements (see Fig.~\ref{atomiclevels}).

Our calculations in Fig.~\ref{LinearKOthersTD} suggest that resolved
structure due to ${^1}\Pi_g$ features may be seen at small detunings
below around $25 \Gamma_{\mrm at}$ for Ca and Ba.  Structure for Sr is
predicted to be suppressed by strong predissociation broadening.  No
${^1}\Pi_g$ structure was reported for detunings larger than around
$50 \Gamma_{\mrm at}$ in Ref.~\cite{Tiemann00}.  Such ${^1}\Pi_g$
structure in Ca$_2$ at these larger detunings may be hard to see due
to masking by the strong ${^1}\Sigma_u$ features.

Figure \ref{LinearKOthersT0} shows our predictions for Ca and Sr
features at extremely low $T = T_D/1000$.  This is in the $s$-wave
limit where the ${^1}\Pi_g$ structure becomes quite sharp, as
discussed in relation to Fig.~\ref{SingleRes} above.  In this domain
sharp ${^1}\Pi_g$ features should be the dominant features in the trap
loss spectrum.  It is noteworthy that this structure is predicted to
persist even to very small detunings on the order of $\Gamma_{\mrm
at}$.  Thus, if the Sr trap loss experiments of
Ref.~\cite{Gallagher99} could be repeated at these low temperatures,
such features might be measurable.  Figure \ref{Width} predicts that
predissociation widths may be large enough for Sr$_2$ ${^1}\Pi_g$
features at even a few $\Gamma_{\mrm at}$ detuning that observed
broadening in the spectra might be able to determine $P_{pe}$ for the
Sr ${^1}\Pi_g$ SC process.  Thus, low temperature measurements provide
for tests of consistency with high temperature measurements.

\section{Conclusions}\label{conclusions}

We have carried out model calculations of the small-detuning
collisional trap loss spectrum of laser cooled Group II species Mg,
Ca, Sr, and Ba and also Yb.  We consider detunings $\Delta$ up to 50
atomic linewidths $\Gamma_{\mrm at}$ to the red of the ${^1}$S$_0 \to
{^1}$P$_1$ laser cooling transition for these species and treat both
inelastic state-changing collisions and radiative loss.  Although our
calculations are only model calculations because the short-range
molecular potentials are not known to sufficient accuracy, we do
incorporate the correct long-range aspects of the potentials and
spectra.  These calculations are intended as a guide for developing
experimental studies on these systems, which have the advantage that
the collisions are not complicated by molecular hyperfine structure.

We consider both the mK range for Doppler cooling on the allowed
${^1}$S$_0 \to {^1}$P$_1$ transition, and $\mu$K range for Doppler
cooling on the ${^1}$S$_0 \to {^3}$P$_1$ intercombination transition.
Collisions in the mK range involve many partial waves, whereas $\mu$K
collisions only involve $s$-wave collisions.  Our quantum mechanical
calculations avoid semiclassical approximations and properly account
for the threshold properties of the collisions.  Our interpretation of
trap loss collision dynamics is based on a factorization of the
overall probability into parts that represent long-range excitation,
propagation to the short-range region, and short-range radiative or
curve crossing processes that lead to loss.  Thus, we can define an
excitation-transfer coefficient $\kappa(\Delta,T)$, which, unlike the
conventional rate coefficient $K(\Delta,T)$, offers a significant
degree of independence from the details of unknown short-range
processes.  Our analysis shows how analytic formulas in the limits of
small or large detuning can be used to interpret the trap loss
spectrum.

The trap loss spectra in all the Group II systems are influenced by
two molecular transitions, the dipole-allowed ${^1}\Sigma_g \to
{^1}\Sigma_u$ transtion and the dipole-forbidden ${^1}\Sigma_g \to
{^1}\Pi_g$ transition.  The latter becomes allowed at long-range
because of retardation corrections to the transition matrix element.
The ${^1}\Sigma_u$ features are structureless at small detuning and
reduced in magnitude due to spontaneous decay of the excited state as
the atoms approach one another on the excited state molecular
potential.  They show broad vibrationally resolved but rotationally
unresolved photoassociation structure as detuning increases away from
atomic resonance.  On the other hand, the ${^1}\Pi_g$ absorption
always dominates at small detuning.  Resolved ${^1}\Pi_g$ vibrational
and rotational photoassociation structure can persist even to small
detuning, and should be especially prominent at very low temperature.
Measurement of the widths of such features could lead to information
about the short-range probability of the state-changing collisions, at
least for the heavier Group II elements.

There are only very limited data with which we can compare our
calculations.  Our model calculations agree within a factor of two
with the measured Sr trap loss rate coefficient at a single detuning.
Photoassociation spectra for Ca only exist for much larger detuning
than we consider here, but suggest that the probability of the
${^1}\Sigma_u$ state changing process may be much smaller than our
model calculations indicate.  The time is right for more detailed and
complete experimental studies on these Group II systems.  Recent
experimental advances in Group II cooling and trapping suggest that
such studies will be forthcoming.  A number of other directions are
also open for continuing experimental and theoretical studies, for
example, trap loss collisions near the ${^1}$S$_0 \to
{^3}$P$_1$ intercombination line, or collisions associated with two
${^1}$P$_1$ atoms or two ${^3}$P atoms.

\acknowledgements

This work has been supported by the Academy of Finland (projects 43336
and 50314), Nordita, NorFA, the Carlsberg Foundation, and the
U.~S.~Office of Naval Research.  We thank Nils Andersen, Alan
Gallagher, Ernst Rasel, Klaus Sengstock, Jan Thomsen and Carl Williams
for discussions, and E. Czuchaj for sending us the new Mg$_2$ results.


\begin{references}
\bibitem{rare} Noble gas atoms are another group of atoms that have
been extensively laser cooled.  The number of collision channels for
these atoms is less than for alkali atoms, but the collisional loss
processes are dominated by Penning ionization.
\bibitem{Flux} The positive definite transfer function $J_{ee}$
represents an incoming flux, including the effect of multiple
vibrations, and can be greater than unity on resonance.
\bibitem{Corrections} An error in the line strength function in our
computer code resulted in magnitudes for $K$ that were too large in
Ref.~\cite{Machholm99}, and the sum over $\l''$ was truncated too
soon.  The qualitative results given in Ref.~\cite{Machholm99} are
correct, however.
\bibitem{GallagherNote} Since two atoms are lost per collision,
Ref.~\cite{Gallagher99} reported the total atom loss rate coefficient
$\beta$, which is $2K$ with our definition of $K$.
\end{references}
\end{document}